\documentclass[10pt,journal]{IEEEtran}
\usepackage[T1]{fontenc}
\usepackage[utf8]{inputenc}
\usepackage{amsmath,amsfonts}
\usepackage{amssymb,amsthm}
\usepackage{algorithmic}
\usepackage{array}
\usepackage[caption=false,font=normalsize,labelfont=sf,textfont=sf]{subfig}
\usepackage{textcomp}
\usepackage{stfloats}
\usepackage{url}
\usepackage{verbatim}
\def\BibTeX{{\rm B\kern-.05em{\sc i\kern-.025em b}\kern-.08em
    T\kern-.1667em\lower.7ex\hbox{E}\kern-.125emX}}
\usepackage{balance}

\usepackage[pdftex]{graphicx}
\usepackage{color}
\usepackage{longtable}
\usepackage{nicefrac}

\usepackage{enumitem}
\usepackage{multicol}

\newcommand{\node}[1]{\mathrm{#1}}

\newcommand{\ds}{\displaystyle}
\newcommand{\ts}{\textstyle}

\newcommand{\Rr}{{\mathbb R}}

\newcommand{\Aa}{\mathbf{A}}

\newcommand{\subind}[2]{\raisebox{-0.5pt}[0mm][0mm]{\scalebox{.55}{$#1,\hspace{-1pt} #2$}}}

\newcommand{\dist}{\mathrm{d}}

\theoremstyle{plain}
\newtheorem{theorem}{Theorem}[section]
\newtheorem{proposition}[theorem]{Proposition}

\theoremstyle{remark}
\newtheorem{remark}[theorem]{Remark}
\newtheorem{property}[theorem]{Property}

\theoremstyle{definition}
\newtheorem{definition}[theorem]{Definition}

\usepackage{tikz}
\usetikzlibrary{calc,fadings,decorations.pathreplacing}
\usetikzlibrary{matrix,arrows}

\def\minus{%
  \setbox0=\hbox{-}%
  \vcenter{%
    \hrule width\wd0 height \the\fontdimen8\textfont3%
  }%
}

\usepackage[ruled,vlined]{algorithm2e}

\begin{document}

\title{ Graph Wedgelets \\[2mm] \normalsize Adaptive Data Compression on Graphs based on Binary Wedge Partitioning Trees and Geometric Wavelets}

\author{Wolfgang Erb
\thanks{Universit{\`a} degli Studi di Padova, Dipartimento di Matematica \newline ''\mbox{Tullio} Levi-Civita'', Padova, Italy, wolfgang.erb@unipd.it. }
}
\markboth{W. Erb, Graph Wedgelets, 24. Nov. 2022}{W. Erb: Graph Wedgelets}

\maketitle

\begin{abstract}
We introduce graph wedgelets - a tool for data compression on graphs based on the representation of signals by piecewise constant functions on adaptively generated binary graph partitionings. The adaptivity of the partitionings, a key ingredient to obtain sparse representations of a graph signal, is realized in terms of recursive wedge splits adapted to the signal. For this, we transfer adaptive partitioning and compression techniques known for 2D images to general graph structures and develop discrete variants of continuous wedgelets and binary space partitionings. We prove that continuous results on best $m$-term approximation with geometric wavelets can be transferred to the discrete graph setting and show that our wedgelet representation of graph signals can be encoded and implemented in a simple way. Finally, we illustrate that this graph-based method can be applied for the compression of images as well. 
\end{abstract}


\IEEEpeerreviewmaketitle

\section{Introduction}

\IEEEPARstart{I}{n} line with the extraordinarily fast growth of stored and transmitted digital information, and the increase in complexity and interdependency of this big data, there is a strong need of novel compression techniques that are able to efficiently compress large data sets on unstructured or semi-structured domains. In many cases, these data sets and their interrelations can be organized in terms of networks or graphs as underlying domains. Adaptive algorithms able to compress data based on its intrinsic content as well as on the topological structure of the surrounding graph environment are therefore of main importance.

Efficient storage of data in image and signal processing depends on how sparsely the data can be represented in terms of suitable chosen dictionaries. The most common representation techniques for graph signals have corresponding counterparts in image processing and comprise, for instance, analogs of the Fourier transform, the wavelet transform, or more general space-frequency decompositions. A general overview about some of these techniques in graph signal processing can be found in \cite{Ortega2018,shuman2013}.   

Focusing on wavelet constructions, there are several approaches that give raise to a wavelet-type multiresolution analysis on graphs. The most prominent works in this direction include diffusion wavelets \cite{Bremer2006,CoifmanMaggioni2006,Szalam2005}, wavelets and vertex-frequency localized frames based on the spectral decomposition of the graph Laplacian \cite{Hammond2011,shuman2012,shuman2016,shuman2020}, graph wavelet filterbanks \cite{NarangOrtega2012}, lifting based wavelets \cite{Jansen2014,ShenOrtega2010}, as well as the construction of wavelets based on partitioning trees \cite{CoifmanGavish2011,GavishNadlerCoifman2010,Murtagh2007}. Interesting for us is mainly the latter approach. This is for two reasons: partitions of the graph vertices can be generated very adaptively and efficiently by a multitude of available graph clustering techniques, as, for instance, $J$-center clustering \cite{Gonzalez1985} or spectral clustering \cite{vonLuxburg2007}; Haar-type wavelets based on hierarchical partitioning trees are easy and cost-efficient to implement, mainly due to the underlying tree structure and the inherent orthogonality of the involved basis functions. Particular construction of Haar wavelets and dictionaries based on hierarchical spectral clustering, $k$-medoids clustering or spanning trees are, for instance, described in \cite{IrionSaito2014,IrionSaito2014b,IrionSaito2017,Sharpnack2013}. More general wavelet-type orthogonal systems on weighted partitioning trees have been studied in \cite{ChuiFilbirMhaskar2015}. Wavelets based on partitioning trees have several applications in machine learning as well, in particular for scattering networks \cite{cheng2016}, and for semi-supervised learning \cite{GavishNadlerCoifman2010}. Further, in \cite{RustamovGuibas2013} improved Haar wavelets for classes of smooth functions have been computed via a deep learning approach. 

The partitioning trees in the works above are solely guided by the topology of the graph and do not take geometric properties of graph signals into account. As
shown in \cite{GavishNadlerCoifman2010}, the particular structure of the partitions has however a strong impact on how well a signal can be approximated sparsely in terms of the Haar wavelets. For an efficient compression of graph signals it is therefore essential that the partitioning trees are adapted to the signal to be compressed.

Goal of this work is therefore to go one step beyond the established non-adaptive constructions of partitioning trees and to develop and analyze new partitioning strategies for graph wavelets that allow for a signal-driven adaptivity in the refinement of the partitions. This can be regarded as an attempt to introduce a new generation of geometric signal-adapted wavelets intrinsically defined on graphs. 

For the compression of images, several approaches for the generation of adaptive partitions are known. Using a function on a continuous 2D domain to describe the image, these approaches usually involve an adaptive segmentation of the image in which the image is approximated by piecewise constant or polynomial functions on the extracted segments. The main idea of this type of compression scheme, and, at the same type also the inherent challenge, is to find a cost-efficient and meaningful splitting procedure such that the resulting segmentation contains only a few relevant elements. On these relevant segments the image is then approximated by simple elementary functions, mainly constant functions or low-order polynomials. If such a meaningful segmentation is found, the resulting compression schemes are highly competitive for low-bit compression \cite{RadhaVetterliLeonardi1996}. Important examples of such adaptive segmentation schemes are adaptive triangulations \cite{CohenDynHechtMirebeau2012,DemaretDynIske2006,DemaretIske2006,DemaretIske2015}, quadtree approximations \cite{LeonardiKunt1985,Samet1985}, tetralets \cite{Krommweh2010}, wedgelets \cite{Donoho1999,Friedrich2007,WakinRomberg2003} or binary space partitioning trees \cite{RadhaLeonardiNaylorVetterli1990,RadhaVetterliLeonardi1996}. The latter two, wedgelets and binary space partitioning trees, will be the most relevant for this work, as their main concepts can be transferred easily to partitions on graphs. In particular, for binary space partitionings, there exists a well-developed theory on the $m$-term approximation with geometric wavelets \cite{DekelLeviatan2003,KaraivanovPetrushev2003} that can be translated directly to the graph setting. In this work, the respective discrete partitionings will be called binary graph partitionings (BGPs). Ideas from continuous wedgelet decompositions and binary space partitionings will further lead to the development of the new discrete graph wedgelets.

\subsection{Main Contributions}
\begin{enumerate}
 \item We provide a theoretical framework for the sparse approximation of graph signals with geometric wavelets defined upon adaptive binary graph partitioning (BGP) trees. This will be done in terms of non-linear $m$-term approximation of functions in discrete Besov-type smoothness classes on graphs. This is an adaption to the discrete graph setting of corresponding continuous results developed in \cite{DekelLeviatan2003,KaraivanovPetrushev2003}.  
 \item We will give a simple and highly efficient novel construction of BGP trees in terms of recursive wedgelet splits on graphs. We will refer to them as binary wedge partitioning (BWP) trees. The BWP trees can be implemented and stored cost-efficiently by an ordered set of graph nodes.
 \item In several experiments, we will study the properties of BWP trees and analyze how well signals on graphs or images can be approximated using adaptive BWPs.
\end{enumerate}

\subsection{Basic terminology on graphs} \label{sec:graphtheory}

\noindent In this work, we consider simple graphs $G=(V,E,\mathbf{A},\mathrm{d})$ with the following structural components:
\begin{enumerate}
\item A set $V=\{\node{v}_1, \ldots, \node{v}_{n}\}$ consisting of $n$ graph vertices. 
\item A set $E \subseteq V \times V$ containing all edges $e_{\subind{i}{i'}} = (\node{v}_i, \node{v}_{i'})$, $i \neq i'$, of the graph $G$. We will assume that $G$ is undirected.  
\item A symmetric adjacency matrix $\Aa \in \Rr^{n \times n}$ with
\begin{equation} \label{eq:generalizedLaplacian}
\ds {\begin{array}{ll}\; \Aa_{\subind{i}{i'}}>0& \text{if $i \neq i'$ and $\node{v}_{i}, \node{v}_{i'}$ are connected,} \\ \; \Aa_{\subind{i}{i'}}=0 & \text{else.}\end{array}}
\end{equation}
The positive elements $\Aa_{\subind{i}{i'}}$, $i \neq i'$, of the adjacency matrix $\Aa$ contain the connection weights of the edges $e_{\subind{i}{i'}} \in E$.

\item The graph geodesic distance $\mathrm{d}$ on the vertex set $V$, i.e., the length of the shortest path connecting two graph nodes. The distance $\mathrm{d}$ satisfies a triangle inequality and, as $G$ is undirected, defines a metric on $V$. We assume that $G$ is a connected graph and, thus, that the distance $\mathrm{d}$ between two arbitrary nodes is finite. 
\end{enumerate}

In this work, we are interested in decompositions of graph signals, i.e. of the functions $x: V \rightarrow \mathbb{R}$ on the vertex set $V$ of the graph $G$. By $\mathcal{L}(V)$, we denote the corresponding $n$-dimensional vector space of graph signals. As the vertices in $V$ are ordered, we can represent every signal $x$ also as a vector $x = (x(\node{v}_1), \ldots, x(\node{v}_n))^{\intercal}\in \mathbb{R}^n$. We can endow the space $\mathcal{L}(V)$ with the inner product 

\begin{equation} y^\intercal x := \sum_{i=1}^n x(\node{v}_i) y(\node{v}_i) \label{eq:innerproductnodes}.
\end{equation} 
The Hilbert space with the norm $\|x\|_{\mathcal{L}^2(V)}^2 = x^\intercal x$, will be denoted as $\mathcal{L}^2(V)$. 
The system $\{\delta_{\node{v}_1}, \ldots, \delta_{\node{v}_n}\}$ of unit vectors forms a canonical orthonormal basis of $\mathcal{L}^2(V)$, where $\delta_{\node{v}_{i'}}$ are defined as $\delta_{\node{v}_{i'}}(\node{v}_i) = \delta_{\subind{i}{i'}}$ for $i,i' \in \{1, \ldots,n\}$. In addition, we consider the $\mathcal{L}^r(V)$ spaces equipped with the quasi-norms
\[\|x\|_{\mathcal{L}^r(V)} = \left( \sum_{i = 1}^n |x(\node{v}_i)|^r \right)^{\frac{1}{r}}, \quad r > 0.\]
It is well-known that for $r \geq 1$ the latter quantity satisfies a triangle inequality and, thus, defines a norm.

\section{Binary graph partitionings (BGPs)}

\noindent The theory of geometric graph wavelets is based on a signal-driven recursive binary partitioning of the vertex set $V$. In particular, the graph partitioning will be adapted to the graph topology as well as on the approximated signal. We start with a general theory on binary partitioning trees on graphs.

\begin{definition} \label{def:BGP} A \emph{binary graph partitioning (BGP) tree} $\mathcal{T}$ of the graph $G$ is a binary tree consisting of subsets of the vertex set $V$ that can be ordered recursively in partitions $\mathcal{P}^{(m)}$, $m \in \mathbb{N}$, of $V$ by the following rules:
\begin{enumerate}
\item The vertex set $V$ is the root of the BGP tree $\mathcal{T}$ and provides the first trivial partition $\mathcal{P}^{(1)} = \{V\}$. 
\item If $\mathcal{P}^{(m)} = \{W_1^{(m)}, \ldots, W_m^{(m)}\}$ is a partition of $V$ consisting of $m$ elements in the BGP tree $\mathcal{T}$, then the next partition $\mathcal{P}^{(m+1)}$ of $V$ in $\mathcal{T}$ is obtained by applying a dyadic split to one of the subsets in $\mathcal{P}^{(m)}$. 
\end{enumerate}
If $W'$ is an element of $\mathcal{P}^{(m+1)}$ obtained from a dyadic split of a set $W \in \mathcal{P}^{(m)}$, then $W' \subset W$ corresponds to a \emph{child} of $W$ in the tree $\mathcal{T}$. We call two elements $W',W'' \in \mathcal{T}$ \emph{siblings} if both are children of the same $W \in \mathcal{T}$. Note that, as $\mathcal{T}$ is binary, a set $W \in \mathcal{T}$ can only have two children or no children at all. In the latter case we call $W$ a \emph{leave} of the tree $\mathcal{T}$. 

We call a BGP tree $\mathcal{T}$ \emph{balanced} if there exists $\frac12 \leq \rho < 1$ such that for every child $W'$ of an element $W \in \mathcal{T}$ we have
\[(1-\rho) |W| \leq |W'| \leq \rho |W|.\]

We call a BGP tree $\mathcal{T}$ \emph{complete} if it has $n$ leaves, each containing a single vertex of the graph $G$. A complete and balanced BGP tree will be referred to as \emph{BGP($\rho$) tree}. 

\end{definition}

To see whether a graph signal $f$ can be approximated sparsely by piecewise constant functions on the elements of a BGP tree, we will analyze the $\mathcal{L}^2$-error $\|f - \mathcal{S}_m (f)\|_{\mathcal{L}^2(V)}$, where $\mathcal{S}_m(f)$ denotes the best $m$-term approximation
\begin{equation} \label{eq:mtermapproximation}
\mathcal{S}_m (f) = \sum_{i = 1}^m \psi_{W_i}(f)(\node{v})
\end{equation}
of the function $f$ in terms of $m$ wavelets $\psi_{W_i}(f)$, $i \in \{1, \ldots, m\}$. These Haar-type wavelets are determined by the elements $W_i$ of a BGP tree $\mathcal{T}$, and sorted descendingly in terms of the $\mathcal{L}^2$-norm: 
\[ \| \psi_{W_1}(f)\|_{\mathcal{L}^2(V)} \geq \| \psi_{W_2}(f)\|_{\mathcal{L}^2(V)} \geq \| \psi_{W_3}(f)\|_{\mathcal{L}^2(V)}  \geq \cdots .\]
The wavelets with respect to a BGP tree $\mathcal{T}$ are defined in the following way: let $W', W \in \mathcal{T}$ such that $W'$ is a child of $W$. Then, the wavelet component $\psi_{W'}(f)$ is given as the signal
\begin{equation} \label{eq:geometricwavelet}
 \psi_{W'}(f)(\node{v}) = \left( \frac{\langle f, \chi_{W'} \rangle}{|W'|} - \frac{\langle f, \chi_{W} \rangle}{|W|}\right) \chi_{W'}(\node{v}),
\end{equation}
where $\chi_{W'}$ denotes the indicator function of the set $W'$. In this way, we obtain for every child $W'$ in $\mathcal{T}$ a wavelet component $\psi_{W'}(f)$ of $f$. For the root $V \in \mathcal{T}$, we additionally set
$$\psi_{V}(f)(\node{v}) = \frac{\langle f, \chi_V \rangle}{|V|}.$$
Now, picking the $m$ components with the largest $\mathcal{L}^2$-norm, we obtain exactly the non-linear $m$-term approximation $\mathcal{S}_m (f)$ of $f$ given in \eqref{eq:mtermapproximation}. If the BGP tree $\mathcal{T} = \mathcal{T}(f)$ depends on $f$ the respective wavelets are called \emph{geometric wavelets}. 

To study the convergence of the $m$-term approximation the following energy functional is of main relevance (see \cite{DeVore1998} for a general overview). It is the discrete counterpart of a corresponding functional given in \cite{DekelLeviatan2003} for binary space partitionings in hypercubes. In wavelet theory, it is usually used in the characterization of Besov spaces and measures in some sense the sparseness of the wavelet representation of a signal $f$. In our case, this sparseness is strongly related to the  partitions given within the BGP tree $\mathcal{T}$.

\begin{definition} \label{def:renergy}
For $0 < r < \infty$, we define the $r$-energy of the wavelet components of $f$ with respect to a BGP tree $\mathcal{T}$ as
\[\mathcal{N}_r(f,\mathcal{T}) = \left( \sum_{W \in \mathcal{T}} 
\|\psi_{W}(f)\|_{\mathcal{L}^2(V)}^r \right)^{\frac{1}{r}}.\]
\end{definition}

We can say the following about the decomposition of $f$ in terms of BGP wavelets. The proof is given in the Appendix.

\begin{theorem} \label{thm:BGPproperties}
Let $G$ be a graph with $n$ nodes, and $\mathcal{T}$ a BGP($\rho$) tree on $G$, i.e., $\mathcal{T}$ is complete and balanced. Then:
\begin{enumerate}
 \item[(i)] The tree $\mathcal{T}$ contains $2n - 1$ elements. 
 \item[(ii)] For every signal $f \in \mathcal{L}(V)$ we have $$f = \sum_{j = 1}^{2n-1} \psi_{W_j}(f), $$
 i.e., $f$ can be decomposed in terms of $2n -1$ wavelets.
 \item[(iii)] For $0 < r < 2$, we have 
 \[ \|f\|_{\mathcal{L}^2(V)} \leq C \mathcal{N}_r(f,\mathcal{T})\]
 with a constant $C > 0$ depending only on $\rho$. 
\end{enumerate}

\end{theorem}

\section{$m$-term approximation error for geometric wavelets on near best BGP trees}

Similar as the $r$-energy functional $\mathcal{N}_r(f,\mathcal{T})$ also the following Besov-type smoothness term quantifies how well a function $f$ can be approximated with piecewise constant functions on the elements of a BGP tree.

\begin{definition} \label{def:besov}
For $\alpha > 0$, $\frac12 \leq \rho < 1$, and $0 < r < \infty$, we define the geometric Besov-type smoothness measure $| \cdot |_{\mathcal{GB}_{r}^{\alpha}}$ of a function $f \in \mathcal{L}(V)$ as
\[| f |_{\mathcal{GB}_{r}^{\alpha}}  \!=  \!\!\inf_{\mathcal{T} \in \mathrm{BGP(\rho)}} \! \left(\sum_{W \in \mathcal{T}} \!\! |W|^{-\alpha r} \! \sup_{\node{w} \in W} \! \sum_{\node{v} \in W} |f(\node{v}) - f(\node{w})|^r \! \right)^{\!\!\frac{1}{r}}.\]
\end{definition}
In \cite{DekelLeviatan2003} (and similarly in \cite{KaraivanovPetrushev2003}), the corresponding spaces of functions have been referred to as geometric B-spaces. In contrast to the $r$-energy introduced in Definition \ref{def:renergy}, the smoothness measure $| f |_{\mathcal{GB}_{r}^{\alpha}}$ is not linked to one particular BGP tree but allows to quantify the sparseness of $f$ with respect to a largy family of BGP$(\rho)$ trees. This can be taken into account also for the $r$-energy $\mathcal{N}_r(f,\mathcal{T})$ by calculating the infimum over all possible BGP$(\rho)$ trees.  

In practice, it might not be possible to determine the infimum over all trees, only an approximate solution might be feasible. Therefore, we say that an $f$-adapted $\mathrm{BGP}(\rho)$ tree $\mathcal{T}_r(f)$ is a near best $\mathrm{BGP}(\rho)$ tree if there exists a constant $C >0$ such that
\[\mathcal{N}_r(f,\mathcal{T}_r(f)) \leq C \inf_{\mathcal{T} \in \mathrm{BGP}(\rho)} \mathcal{N}_r(f,\mathcal{T}).\]
This is the setting we have in mind when we design greedy algorithms in the next section to create adaptive partitionings for the compression of graph signals. For near best $\mathrm{BGP}(\rho)$ trees, we have the following relation. 

\begin{theorem} \label{thm:estimate}
Let $\alpha > 0$, $\frac12 \leq \rho < 1$ and $1/r = \alpha + 1/2$. Further, let $\mathcal{T}_r(f)$ be a near best $\mathrm{BGP}(\rho)$ tree. Then, we have the equivalences
\[C_1 \mathcal{N}_r(f,\mathcal{T}_r(f)) \leq |f|_{\mathcal{GB}_{r}^{\alpha}} \leq C_2 \mathcal{N}_r(f,\mathcal{T}_r(f))\]
with constants $C_1$ and $C_2$ that depend only on $\alpha$ and $\rho$. 
\end{theorem}

We can finally conclude that if $f$ is smooth with respect to the Besov measure given in Definition \ref{def:besov}, it suffices to find a near best $\mathrm{BGP}(\rho)$ tree to obtain the following $m$-term approximation rates. 

\begin{theorem}[Jackson estimate] \label{thm:Jackson}
Let $\alpha > 0$ and $r > 0$ be related by $1/r = \alpha + 1/2$. Then, for a graph signal $f \in \mathcal{L}(V)$ and geometric wavelets with respect to a near best BGP$(\rho)$ tree $\mathcal{T}_r(f)$, we obtain the $m$-term approximation error
\[ \left\| f - \mathcal{S}_m(f) \right\|_{\mathcal{L}^2(V)} \leq C m^{-\alpha} |f|_{\mathcal{GB}_{r}^{\alpha}} \]
with a constant $C>0$ that depends only on $r$ and $\rho$.  
\end{theorem}

Both, Theorem \ref{thm:estimate} and \ref{thm:Jackson} are discrete versions of respective continuous results given for geometric wavelets on binary space partitionings \cite{DekelLeviatan2003} and piecewise polynomial approximation on nested triangulations \cite{KaraivanovPetrushev2003}. The proofs are provided in the appendix of this article. We note that all the results of this work hold true also more generally for discrete metric spaces. Within this class, graphs equipped with the graph geodesic distance as a metric are however the most relevant examples for us.

\section{BWP trees and graph wedgelets }

\begin{figure}[htbp]
	\centering
	\includegraphics[width=0.44\textwidth]{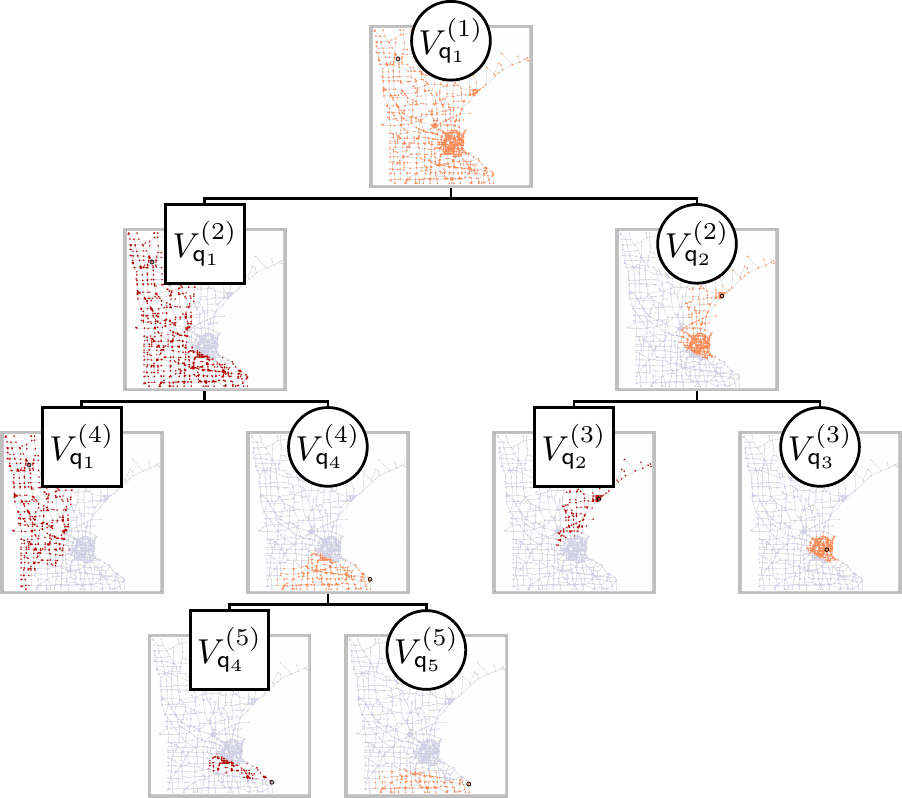}  
	\caption{A BWP tree for the adaptive approximation of the test function $f_2$ in Fig. \ref{fig:ringeling} (middle) on the Minnesota graph.}
	\label{fig:BWPtree-minnesota}
\end{figure}

\noindent After the general theory on BGPs, we are interested in finding explicit constructions of adaptive partitioning trees on graphs. For an efficient coding of the BGP trees we require simple splitting strategies on graphs that can be adapted to the geometry of a graph signal. For this, we propose as building elements the following elementary wedge splits. 

\begin{definition} \label{def:wedgesplit}
We call a dyadic partition $\{V', V''\}$ of the vertex set $V$ a \emph{wedge split} of $V$ if there
exist two distinct nodes $\node{v}'$ and $\node{v}''$ of $V$ such that $V'$ and $V''$ have the form
\begin{align*} V' &= \{\node{v} \in V \ | \ \dist(\node{v},\node{v}') \leq \dist(\node{v},\node{v}'')\}, \quad \text{and} \\  V'' &= \{\node{v} \in V \ | \ \dist(\node{v},\node{v}') > \dist(\node{v},\node{v}'')\}.
\end{align*} 
\end{definition}

\noindent A key advantage of the just defined wedge splits is that they can be encoded very compactly in terms of the two nodes $\node{v}'$ in $ V'$ and $\node{v}''$ in $V''$. They have the following basic properties:

\begin{property}
If $\{V', V''\}$ is a wedge split of $V$, then
\begin{enumerate}
\item $V'$ and $V''$ are uniquely determined by $\node{v}'$ and $\node{v}''$. 
\item $V' \cap V'' = \emptyset$ and $V' \cup V'' = V$.
\item If the vertex set $V$ is connected, then also $V'$ and $V''$ are connected subsets of the graph $G$. 
\end{enumerate}
\end{property}

While the first two properties follow immediately from the Definition \ref{def:wedgesplit}, the third property is a consequence of the fact that $\mathrm{d}$ is the shortest path distance on $G$. 

Using wedge splits we can define elementary wedgelets $\omega_{(\node{v}',\node{v}'')}^{+}$ and $\omega_{(\node{v}',\node{v}'')}^{-}$ as the following indicator functions:
\begin{align*} \omega_{(\node{v}',\node{v}'')}^{+} (\node{v}) &= \chi_{V'}(\node{v}) = \left\{\begin{array}{ll} 1, & \text{if $\dist(\node{v},\node{v}') \leq \dist(\node{v},\node{v}'')$}, \\
0, & \text{otherwise}, \end{array} \right. \\
\omega_{(\node{v}',\node{v}'')}^{-} (\node{v}) &= \chi_{V''}(\node{v}) = \left\{\begin{array}{ll} 1, & \text{if $\dist(\node{v},\node{v}') > \dist(\node{v},\node{v}'')$}, \\
0, & \text{otherwise}. \end{array} \right. 
\end{align*}

\noindent Using wedge splits, we generate now the following BGP trees:   

\begin{definition} \label{def:BWP} A binary wedge partitioning (BWP) tree $\mathcal{T}_Q$ of $G$ with respect to the ordered set $Q = \{\node{q}_1, \ldots, \node{q}_M\} \subset V$ is a BGP tree constructed recursively as follows:
\begin{enumerate}
\item The root of $\mathcal{T}_Q$ is the set $V$. It forms the trivial partition $\mathcal{P}^{(1)} = \{V_{\node{q}_1}^{(1)}\} = \{V\}$ and is associated to $\node{q}_1 \in Q$. 
\item For a partition $\mathcal{P}^{(m)} = \{ V_{\node{q}_1}^{(m)}, \ldots, V_{\node{q}_m}^{(m)} \}$ of $V$ in $\mathcal{T}_Q$ associated to $\node{q}_i \in V_{\node{q}_i}^{(m)}$, $i \in \{1, \ldots, m\}$, $m < M$, consider the node $\node{q}_{m+1} \in V_{\node{q}_j}^{(m)}$ for a $j \in \{1, \ldots, m\}$. We split $V_{\node{q}_j}^{(m)}$ by a wedge split based on $\node{q}_{j}$ and $\node{q}_{m+1}$ into two disjoint sets $V_{(\node{q}_j,\node{q}_{m+1})}^{(m) \, +}$ (containing $\node{q}_j$) and $V_{(\node{q}_j,\node{q}_{m+1})}^{(m) \, -}$ (containing $\node{q}_{m+1}$) and obtain the new partition $$ \mathcal{P}^{(m+1)} = \{ V_{\node{q}_1}^{(m+1)}, \ldots, V_{\node{q}_{m+1}}^{(m+1)} \}$$ with $V_{\node{q}_i}^{(m+1)} = V_{\node{q}_i}^{(m)}$ if $i \neq \{j,m+1\}$, $V_{\node{q}_j}^{(m+1)} = V_{(\node{q}_j,\node{q}_{m+1})}^{(m) \, +}$ and $V_{\node{q}_{m+1}}^{(m+1)} =V_{(\node{q}_j,\node{q}_{m+1})}^{(m) \, -}$.  
\end{enumerate}
\end{definition}

A BWP tree $\mathcal{T}_Q$ as given in Definition \ref{def:BWP} is uniquely determined by the ordered  node set $Q$. This allows to store $\mathcal{T}_Q$ compactly in terms of the $M$ nodes of $Q$.  Further, adaptive refinements of a BWP tree can be coded directly in terms of the centers $Q$, i.e., in every refinement step an adaptive selection of a node $\node{q}_{j} \in Q$ and the generation of new node $\node{q}_{m+1}$ is necessary. We collect some properties of BWP trees that follow directly from Definition \ref{def:BWP}, Theorem \ref{thm:BGPproperties} and the general Definition \ref{def:BGP} for BGPs.

\begin{proposition} \label{prop:BWP}
Let $\mathcal{T}_Q$ be a BWP tree determined by the ordered node set $Q = \{\node{q}_1, \ldots, \node{q}_M\}$. 
\begin{enumerate}
\item A BWP tree $\mathcal{T}_Q$ contains $2M -1$ elements: $1$ root and $2M - 2$ children. 
\item The $M$ leaves of the binary tree $\mathcal{T}_Q$ are given by the elements of the $M$-th. partition $\mathcal{P}^{(M)} = \{ V_{\node{q}_1}^{(M)}, \ldots, V_{\node{q}_{M}}^{(M)}\}$.  
\item A BWP tree $\mathcal{T}_Q$ is complete if and only if $|Q| = |V|$.  
\item A BWP tree $\mathcal{T}_Q$ is balanced with
$\frac12 \leq \rho \leq \frac{n-1}{n}$.
\item The characteristic function of the subset $V_{\node{q}_i}^{(m)}$ can be written as a product of $m$ elementary wedgelets $\omega_{(\node{q}_i,\node{q}_j)}^{\pm}$, with $\node{q}_i, 
\node{q}_j \in \{\node{q}_1, \ldots, \node{q}_m\}$, $i < j$.

\end{enumerate}
\end{proposition}

\begin{definition} \label{def:wedgelets}
The characteristic functions 
\[\omega_{\node{q}_i}^{(m)}(\node{v}) = \chi_{V_{\node{q}_i}^{(m)}}(\node{v}), \quad 1 \leq i \leq m, \; 1 \leq m \leq M,\]
of the sets $V_{\node{q}_i}^{(m)}$ will be referred to as \emph{wedgelets} with respect to the BWP tree $\mathcal{T}_Q$.
The wedgelets $\{\omega_{\node{q}_i}^{(m)}\,:\,1 \leq i \leq m\}$ form an orthogonal basis for the piecewise constant functions on the partition $\mathcal{P}^{(m)}$ (using the inner product \eqref{eq:innerproductnodes} in $\mathcal{L}^2(V)$).
\end{definition}

\begin{remark}
It is possible to define larger families of BWP trees than in Definition \ref{def:BWP} (giving larger sets of wedgelets in Definition \ref{def:wedgelets}, respectively) by allowing wedge splits inside the subsets  that are not linked to the centers $\node{q}_i$. In this case, the tree can however not be represented with a simple node set $Q$. It leads also to a larger computational cost in the selection of a proper wedge split when calculating the adaptive tree.
\end{remark}

\section{Adaptive greedy generation of BWP trees}
\noindent To generate BWP trees $\mathcal{T}_Q$ that are adapted to a given graph signal $f$, every refinement step requires two principal pieces of information. At each partition level $m$ one of the sets $V_{\node{q}_j}^{(m)}$, $j \in \{1, \ldots, m\}$, has to be selected, and a new node $\node{q}_{m+1} \in V_{\node{q}_j}^{(m)}$ is required for the consequent elementary wedge split of $V_{\node{q}_j}^{(m)}$. 
In general, both choices can be made in an $f$-adapted or in a non-adaptive way. As adaptive refinement procedures, we consider the following three greedy methods:

{\noindent \bfseries Max-distance (MD) greedy wedge splitting:} at stage $m$, the domain $V_{\node{q}_j}^{(m)}$ is chosen $f$-adaptively by selecting $j$ such that
\begin{equation} \label{eq:greedyset}
j = \underset{i \in \{1, \ldots, m\}}{\mathrm{argmax}} \|f - \bar{f}_{V_{\node{q}_i}^{(m)}}\|_{\mathcal{L}^2(V_{\node{q}_i}^{(m)})},\end{equation}
where 
\[ \textstyle \bar{f}_{V_{\node{q}_i}^{(m)}} = \frac{\langle f, \omega_{\node{q}_i}^{(m)} \rangle}{|V_{\node{q}_i}^{(m)}|} = \frac{1}{|V_{\node{q}_i}^{(m)}|} \underset{\node{v} \in V_{\node{q}_i}^{(m)}}{\sum} f(\node{v})\]
denotes the mean value of $f$ over the set $V_{\node{q}_i}^{(m)}$. This first selection rule ensures that the chosen set $V_{\node{q}_j}^{(m)}$ is the one with the maximal $\mathcal{L}^2$-error in \eqref{eq:greedyset}. As soon as $j$, or equivalently, $\node{q}_j$ are determined, a non-adaptive way to choose the subsequent node set $\node{q}_{m+1}$ is by the selection rule
\[ \node{q}_{m+1} = \mathrm{arg\, max}_{\node{v} \in V_{\node{q}_j}^{(m)}} \, \mathrm{d}(\node{q}_j,\node{v}),
\]
i.e., $\node{q}_{m+1}$ is the vertex in $V_{\node{q}_j}^{(m)}$ furthest away from $\node{q}_j$. This choice and the corresponding split can be interpreted as a two center clustering of $V_{\node{q}_j}^{(m)}$ in which the first node $\node{q}_j$ is fixed (see a previous work \cite{cavoretto2021} for more details on greedy $J$-center clustering). One heuristic reason for this selection is that the resulting binary partitions in the BWP tree might be more balanced with a smaller constant $\rho$ compared to the theoretical upper bound $1 - 1/n$ in Proposition \ref{prop:BWP}. 

{\noindent \bfseries Fully-adaptive (FA) greedy wedge splitting:} in the FA-greedy procedure the subset to be split is selected according to \eqref{eq:greedyset}, but also the node $\node{q}_{m+1}$ determining the wedge split is chosen according to an adaptive rule. If $\textstyle \{ V_{(\node{q}_j,\node{q})}^{(m) \, +}, V_{(\node{q}_j,\node{q})}^{(m) \, -}\}$ denotes the partition of $V_{\node{q}_j}^{(m)}$ for the wedge split determined by $\node{q}_j$ and a second node $\node{q}$, we choose $\node{q}_{m+1}$ such that
\begin{equation} \label{eq:fullyadaptivegreedy}  \|f - \bar{f}_{V_{(\node{q}_j,\node{q})}^{(m) \, +}}\|_{\mathcal{L}^2(V_{(\node{q}_j,\node{q})}^{(m) \, +})}^2 + \|f - \bar{f}_{V_{(\node{q}_j,\node{q})}^{(m) \, -}}\|_{\mathcal{L}^2(V_{(\node{q}_j,\node{q})}^{(m) \, -})}^2
\end{equation}
is minimized over all $\node{q} \in V_{\node{q}_j}^{(m)}$. Compared to the semi-adaptive MD-greedy procedure, the FA-greedy method is computationally more expensive. On the other hand, as the wedge splits are more adapted to the particular form of the underlying function $f$, we expect a better approximation behavior for the FA-greedy scheme. This expectation will be confirmed in the numerical experiments performed in the last section. 

{\noindent \bfseries Randomized (R) greedy wedge splitting:} If the size of the subsets $V_{\node{q}_j}^{(m)}$ is large it might be too time-consuming to find the global minimum of the quantity \eqref{eq:fullyadaptivegreedy} in the FA-greedy scheme. A quasi-optimal alternative to the fully-adaptive procedure is a randomized splitting strategy, in which the minimization of \eqref{eq:fullyadaptivegreedy} is performed on a subset of $1 \leq R \leq |V_{\node{q}_j}^{(m)}|$ randomly picked nodes of $V_{\node{q}_j}^{(m)}$. In this strategy, the parameter $R$ acts as a control parameter giving a result close or identical to FA-greedy if $R$ is chosen large enough. 

The just described adaptive selection rules to generate BWP trees and the respective wedgelet encoding and decoding variants are summarized in Algorithm \ref{alg:wedgeletencoding} and Algorithm \ref{alg:wedgeletdecoding}. 

\begin{algorithm}  
\small

\caption{Wedgelet encoding of a graph signal}

\label{alg:wedgeletencoding}

\vspace{1mm}

\KwIn{Graph signal $f$, initial node $\node{q}_1 \in V$, first partition $\mathcal{P}^{(1)} = \{V\} = \{V_{\node{q}_1}^{(1)}\}$ and final partition size $M$.  
}

\vspace{1mm}

\For{$m = 2$ to $M$}
    {1) {\bfseries Greedy selection of subset:} calculate $j$ according to the rule \eqref{eq:greedyset} as
    \[
    j = \underset{i \in \{1, \ldots, m-1\}}{\mathrm{arg \, max}} \big\|f - \bar{f}_{V_{\node{q}_i}^{(m-1)}} \big\|_{\mathcal{L}^2(V_{\node{q}_i}^{(m-1)})};\]
    
    2) Conduct one of the following alternatives:
    
    {\bfseries Max-distance (MD) greedy procedure:} select new node $\node{q}_m = \mathrm{arg \, max}_{\node{v} \in V_{\node{q}_j}^{(m)}} \; \dist(\node{q}_j,\node{v})$
    farthest away from $\node{q}_j$ and add it to the node set $Q$;
        
    {\bfseries Fully-adaptive (FA) greedy procedure:} determine new node $\node{q}_m$ such that the squared $\mathcal{L}^2$-error term \eqref{eq:fullyadaptivegreedy} is minimized and add it to the node set $Q$; 
    
    {\bfseries Randomized (R) greedy procedure:} determine $\node{q}_m$ such that \eqref{eq:fullyadaptivegreedy} is minimized over a subset of $R$ randomly chosen points and add it to $Q$;
    
    3) According to Definition \ref{def:BWP}, generate the {\bfseries new partition} $\mathcal{P}^{(m)}$ from the partition $\mathcal{P}^{(m-1)}$ by a wedge split of the subset $V_{\node{q}_j}^{(m-1)}$ into the children sets $V_{(\node{q}_j,\node{q}_{m})}^{(m-1) \, +}$ and $V_{(\node{q}_j,\node{q}_m)}^{(m-1) \, -}$;
    
    4) Compute {\bfseries mean values} $\bar{f}_{V_{\node{q}_i}^{(m)}}$, $i \in \{1, \ldots, m\}$, for the
    new partition $\mathcal{P}^{(m)}$ by an update from $\mathcal{P}^{(m-1)}$.}
 
\vspace{1mm}    

\KwOut{$Q = \{\node{q}_1, \ldots, \node{q}_{M}\}$, $\big\{\bar{f}_{V_{\node{q}_1}^{(M)}}, \ldots, \bar{f}_{V_{\node{q}_M}^{(M)}}\big\}$.}

\end{algorithm}

\begin{algorithm} 
\small

\caption{Wedgelet decoding of a graph signal}

\label{alg:wedgeletdecoding}

\vspace{1mm}

\KwIn{$Q = \{\node{q}_1, \ldots, \node{q}_{M}\}$, $\big\{\bar{f}_{V_{\node{q}_1}^{(M)}}, \ldots, \bar{f}_{V_{\node{q}_M}^{(M)}}\big\}$.  
}

\vspace{2mm}

{\bfseries Calculate} the partition $\mathcal{P}^{(M)} = \{V_{\node{q}_1}^{(M)}, \ldots, V_{\node{q}_M}^{(M)}\}$ of $V$ by elementary wedge splits along the BWP tree $\mathcal{T}_Q$ according to the recursive procedure in Definition \ref{def:BWP}.  
 
\vspace{2mm}    

\KwOut{The wedgelet approximation
\[\mathcal{W}_M f(\node{v}) = \sum_{i = 1}^M \bar{f}_{V_{\node{q}_i}^{(M)}} \, \omega_{\node{q}_i}^{(M)} (\node{v})\]
of $f$. For $M = n$, $\mathcal{W}_n f = f$ is reconstructed.}

\end{algorithm}

{\noindent \bfseries Upper bounds for the computational cost:}
We assume that the distance between two nodes can be calculated in $\mathcal{O}(1)$ operations. Further, in a worst-case scenario, the calculation of a mean value over $V_{\node{q}_j}^{(m)}$ requires $\mathcal{O}(n)$ operations. The computational expenses of Algorithm \ref{alg:wedgeletencoding} can thus be bounded by $\mathcal{O}(M n^2)$, $\mathcal{O}(M R n)$ and $\mathcal{O}(M n)$ operations for FA-greedy, R-greedy and MD-greedy, respectively. In Fig. \ref{fig:BWP-church} e), the computational times required by the three BWP variants are plotted for an image decomposition. The measured times indicate that the dependence of the cost on $M$ is rather sublinear than linear.

{\noindent \bfseries Acceleration possibilities:}
For very large $n$, FA-greedy and R-greedy (for large $R$) might be too expensive. In this case, a possibility to increase the calculational speed is to split the graph a priori into $J$ subgraphs. Then, the adaptive BWP methods can be applied (also in a parallelized form) separately to each subgraph. For this procedure, clustering algorithms as $J$-center clustering \cite{Gonzalez1985} are available. In \cite{cavoretto2021}, this clustering method has been used for partition of unity methods on graphs.  

\subsection{Geometric wavelets based on wedge splits}

Instead of storing the mean values $\big\{\bar{f}_{V_{\node{q}_1}^{(M)}}, \ldots, \bar{f}_{V_{\node{q}_M}^{(M)}}\big\}$ of the wedgelet approximation $\mathcal{W}_M f$, we can alternatively encode $\mathcal{W}_M f$ using the geometric wavelets introduced in \eqref{eq:geometricwavelet}. This alternative representation is particularly suited if a further compression of the signal is desired, for instance by using an $m$-term approximation of the signal with $m < M$.

For a BWP tree $\mathcal{T}_Q$ and $2 \leq m \leq M$, we define the index $j = j(m) < m$ such that $V_{\node{q}_{j}}^{(m)}$ is the sibling of $V_{\node{q}_{m}}^{(m)}$ in the BWP tree $\mathcal{T}_Q$. Then, the geometric wavelets with respect to the BWP tree $\mathcal{T}_Q$ are defined as the signals
\begin{align*}
 \psi_{(\node{q}_j,\node{q}_m)}^+(f) &= \left(\! \bar{f}_{V_{\node{q}_{j}}^{(m)}} -\bar{f}_{V_{\node{q}_{j}}^{(m-1)}} \! \right) \omega_{\node{q}_{j}}^{(m)} =: c_{(\node{q}_j,\node{q}_m)}^+(f) \, \omega_{\node{q}_{j}}^{(m)}, \\
  \psi_{(\node{q}_j,\node{q}_m)}^-(f) &= \left(\! \bar{f}_{V_{\node{q}_{m}}^{(m)}} - \bar{f}_{V_{\node{q}_{j}}^{(m-1)}} \! \right) \omega_{\node{q}_{m}}^{(m)} =: c_{(\node{q}_j,\node{q}_m)}^-(f) \, \omega_{\node{q}_{m}}^{(m)}.
\end{align*}
Further, in the particular case $m = 1$, we set $$\psi_{\node{q}_1}(f) = \bar{f}_{V} \, \omega_{\node{q}_{1}}^{(1)} =: c_{\node{q}_1}(f)\,\omega_{\node{q}_{1}}^{(1)}.$$ In this way, we obtain $2M-1$ geometric wavelets for $\mathcal{T}_Q$. Beside the altered notation, this definition corresponds to the general definition of geometric wavelets for BGP trees given in \eqref{eq:geometricwavelet}. 
For $\psi_{(\node{q}_j,\node{q}_m)}^+(f)$ and $\psi_{(\node{q}_j,\node{q}_m)}^-(f)$, we have the relation
\begin{align*}
\sum_{\node{v} \in V} & \left( \psi_{(\node{q}_j,\node{q}_m)}^+(f)(\node{v}) + \psi_{(\node{q}_j,\node{q}_m)}^-(f)(\node{v}) \right) \\& =
\bar{f}_{V_{\node{q}_{m}}^{(m)}} |V_{\node{q}_{m}}^{(m)}| + \bar{f}_{V_{\node{q}_{j}}^{(m)}} |V_{\node{q}_{j}}^{(m)}| - \bar{f}_{V_{\node{q}_{j}}^{(m-1)}} |V_{\node{q}_{j}}^{(m-1)}| \\
&=\sum_{\node{v} \in V_{\node{q}_{m}}^{(m)}} f(\node{v}) + \sum_{\node{v} \in V_{\node{q}_{j}}^{(m)}} f(\node{v}) -  \sum_{\node{v} \in V_{\node{q}_{j}}^{(m-1)}} f(\node{v}) = 0.
\end{align*}
This implies that $\langle \psi_{(\node{q}_j,\node{q}_m)}^+(f), 1 \rangle = - \langle \psi_{(\node{q}_j,\node{q}_m)}^-(f), 1 \rangle$
and that $c_{(\node{q}_j,\node{q}_m)}^+(f)$ and $c_{(\node{q}_j,\node{q}_m)}^-(f)$ are related by 
\[c_{(\node{q}_j,\node{q}_m)}^+(f) = \textstyle - \frac{|V_{\node{q}_{m}}^{(m)}|}{|V_{\node{q}_{j}}^{(m)}|}c_{(\node{q}_j,\node{q}_m)}^-(f).\]
In particular, when computing the coefficients of the geometric wavelets, it suffices to store one of the two, either $c_{(\node{q}_j,\node{q}_m)}^+(f)$ or $c_{(\node{q}_j,\node{q}_m)}^-(f)$. The encoding and decoding of a graph signal $f$ in terms of wedge-based geometric wavelets is compactly described in Algorithm \ref{alg:waveletencoding} and Algorithm \ref{alg:waveletdecoding}.

\begin{algorithm} 

\small

\caption{Wedgelet encoding with BWP wavelets}

\label{alg:waveletencoding}

\vspace{1mm}

\KwIn{Graph signal $f$, initial node $\node{q}_1 \in V$, first partition $\mathcal{P}^{(1)} = \{V\} = \{V_{\node{q}_1}^{(1)}\}$ and final partition size $M$.  
}

\vspace{1mm}

\For{$m = 2$ to $M$}
    {1) 2) \& 3) as in Algorithm 1;
    
    4) Compute {\bfseries geometric wavelet coefficients} $c_{(\node{q}_j,\node{q}_m)}^+$ and $c_{(\node{q}_j,\node{q}_m)}^-$ (one of the two is sufficient).}
 
\vspace{1mm}    

\KwOut{$Q = \{\node{q}_1, \ldots, \node{q}_{M}\}$, $\big\{c_{\node{q}_1}, c_{(\node{q}_j,\node{q}_2)}^\pm, \ldots, c_{(\node{q}_j,\node{q}_M)}^\pm\big\}$.}

\end{algorithm}

\begin{algorithm} 
\small

\caption{Wedgelet decoding with BWP wavelets}

\label{alg:waveletdecoding}

\vspace{1mm}

\KwIn{$Q = \{\node{q}_1, \ldots, \node{q}_{M}\}$, $\big\{c_{\node{q}_1}, c_{(\node{q}_j,\node{q}_2)}^\pm, \ldots, c_{(\node{q}_j,\node{q}_M)}^\pm\big\}$.

}

\vspace{2mm}

$\mathcal{W}_1 f = c_{\node{q}_1}$.  

\vspace{2mm}

\For{$m = 2$ to $M$}
{{\bfseries Calculate} the partition $\mathcal{P}^{(m)} = \{V_{\node{q}_1}^{(m)}, \ldots, V_{\node{q}_m}^{(m)}\}$ of $V$ from the partition $\mathcal{P}^{(m-1)}$ by an elementary wedge split of the set $V_{\node{q}_j}^{(m-1)}$.
{\bfseries Update} the wedgelet approximation
\[\mathcal{W}_m f =  \mathcal{W}_{m-1} f + c_{(\node{q}_j,\node{q}_m)}^+ \omega_{\node{q}_{j}}^{(m)} + c_{(\node{q}_j,\node{q}_m)}^- \omega_{\node{q}_{m}}^{(m)}.\]
}

\vspace{2mm}    

\KwOut{The wedgelet approximation
\[\mathcal{W}_M f = c_{\node{q}_1} + \sum_{m = 2}^M \left(c_{(\node{q}_j,\node{q}_m)}^+ \omega_{\node{q}_{j}}^{(m)} + c_{(\node{q}_j,\node{q}_m)}^- \omega_{\node{q}_{m}}^{(m)}\right)\]
of $f$. For $M = n$, $\mathcal{W}_n f = f$ is reconstructed.}

\end{algorithm}

\section{Memory requirements for wedgelet encoding} \label{sec:storage}
To encode a graph signal in terms of $M$ wedgelets, we need to store the mean values $\big\{\bar{f}_{V_{\node{q}_1}^{(M)}}, \ldots, \bar{f}_{V_{\node{q}_M}^{(M)}}\big\}$ on the leaves of the BWP tree as well as the geometric information of the wedge splits provided by the set $Q = \{\node{q}_1, \ldots, \node{q}_{M}\}$. Using a quantization of $K$ different values, every mean value $\bar{f}_{V_{\node{q}_i}^{(M)}}$ can be stored with $\log_2(K)$ bits. The same holds true if the BWP wavelet coefficients $c_{(\node{q}_j,\node{q}_m)}^+$ replace the mean values. As we have $n$ nodes, we can further store any node $\node{v}_j$ using its index $1\leq j \leq n$. This requires at most $\log_2(n) M$ bits to store the node set $Q$.  Consequently, we get the following upper bound for the memory requirements of wedgelet encoding. 

\begin{theorem} \label{thm:memory}
Assume that the mean values $\bar{f}_{V_{\node{q}_i}^{(M)}}$ (or the coefficients $c_{\node{q}_1}$ and $c_{(\node{q}_j,\node{q}_i)}^+$) are given in a quantized form with at most $K$ different values. Then, the wedgelet encodings in Algorithm \ref{alg:wedgeletencoding} and Algorithm \ref{alg:waveletencoding} require a memory of at most \[\frac{\lceil\log_2(n) + \log_2(K)\rceil M}{n} \quad \text{bits per node.}\]
\end{theorem}

In the particular case of an image with $512 \times 512 = 2^{18}$ pixels and an image depth of $K = 2^8 = 256$ colors we get by Theorem \ref{thm:memory} that a representation with $M = 1000$ wedgelets requires a memory of less than $0.1$ bits per pixel.

\section{Examples on graphs and images}

\subsection{Examples of BWPs on graphs}

\begin{figure}[htbp]
	\centering
	\includegraphics[width=0.495\textwidth]{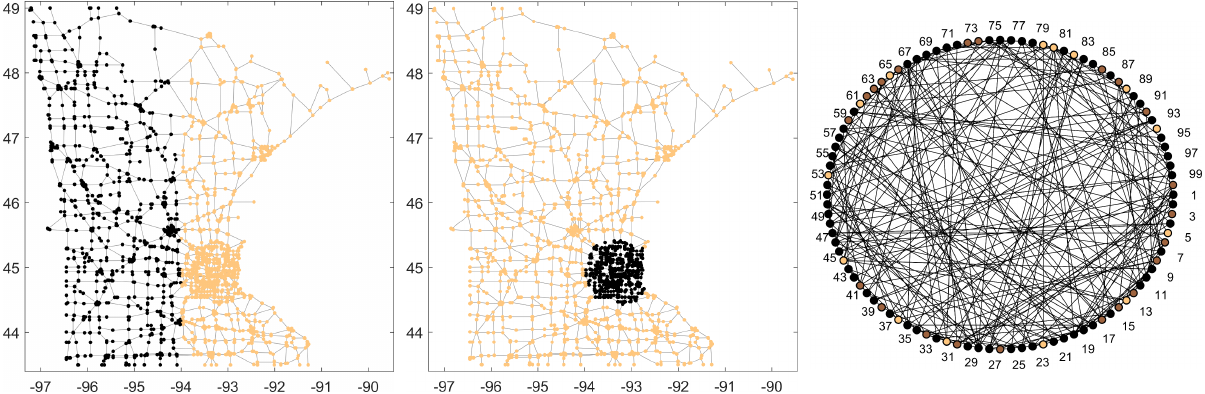}
	\caption{Test signals $f_1$ (left), $f_2$ (middle) and $f_3$ (right) to generate BWPs.}
	\label{fig:ringeling}
\end{figure}

We will consider two test graphs, the Minnesota road network $G_1$ and an Erdős–Rényi graph $G_2$. The dataset for $G_1$ has been retrieved from \cite{RossiAhmed2015} and consists of $n=2642$ vertices and $3304$ edges. The graph $G_2$ with $n = 100$ vertices and $236$ edges was generated according to the $G(n,p)$ Erdős–Rényi random graph model with probability $p = 0.05$. The distance metric on both graphs is the shortest-path distance. As test signals on $G_1$ we consider the two binary functions
\[f_1 = 2 \chi_{V_1} -1, \quad f_2 = 2 \chi_{V_2} -1,\]
based on the characteristic functions of the node sets
\begin{align*} V_1 &= \{\node{v} \in V \ | \ x_\node{v} <-94\}, \\ V_2 &= \{ \node{v} \in V  \ | \ 0.75 (x_\node{v} + 93.3)^2 + (y_\node{v}  -44.95)^2  < 0.35\},
\end{align*}
where $(x_{\node{v}},y_{\node{v}})$ denote the Cartesian coordinates of the node $\node{v} \in V$. The test signal $f_3$ on $G_2$ describes a clustering of $G_2$ based on three integer values from $1$ to $3$. The functions $f_1$, $f_2$ and $f_3$ are illustrated in Fig. \ref{fig:ringeling}. 
 
\begin{figure*}[htbp]
	\centering
	\includegraphics[width=0.95\textwidth]{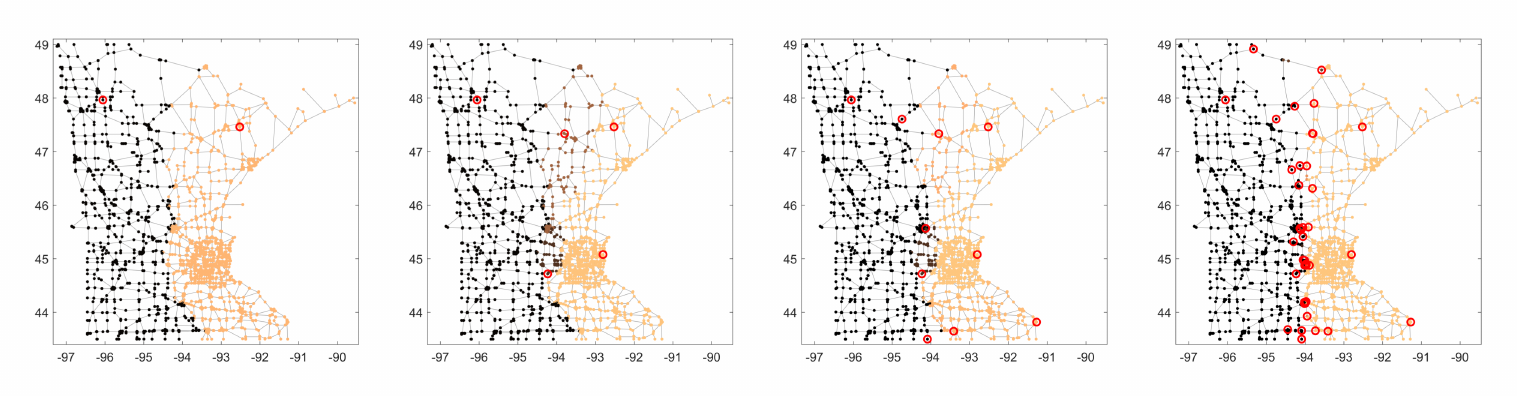} 
	\vspace{-4mm}
	\caption{Approximation of $f_1$ with $1,4,9$ and $39$ wedge splits (from left to right). The red rings indicate the center nodes $Q$. The number of wrongly classified nodes equals $356$, $286$, $110$, and $12$, respectively.}
	\label{fig:BWP-minnesota}
\end{figure*}

Starting from a random node $\node{q}_1$ we use Algorithm \ref{alg:wedgeletencoding} to generate the BWP tree as well as a piecewise constant approximation of the functions. 
The initial part of the BWP tree for the approximation of $f_2$ (using FA-greedy) is shown in Fig. \ref{fig:BWPtree-minnesota}. In Fig. \ref{fig:BWP-minnesota}, approximations $\mathcal{W}_m f_1$ of $f_1$ are illustrated for different partitioning stages $m$ (the number of wedge splits equals $m-1$, we used FA-greedy to generate the BWP tree). 

\begin{figure}[htbp] 
	\centering
	\includegraphics[width=0.495\textwidth]{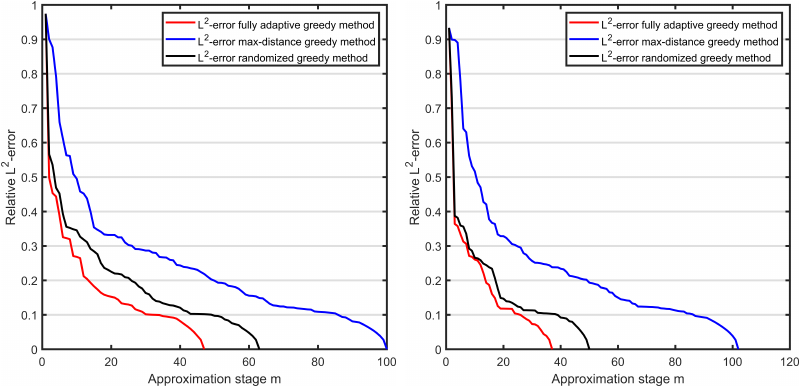} 	
	\vspace{-4mm}
	\caption{ $L^2$-approximation errors for the BWP approximation $\mathcal{W}_m f$ of $f_1$ (left) and $f_2$ (right) using FA-greedy, MD-greedy and R-greedy with $R = 50$.}
	
\label{fig:semivsfullminnesota}
\end{figure}

The three variants MD-greedy, R-greedy and FA-greedy for the test functions $f_1$ and $f_2$ are compared in Fig. \ref{fig:semivsfullminnesota}. In these examples, FA-greedy performs best for most partitioning stages $m$, followed by R-greedy and MD-greedy. The FA-greedy variant is also the most cost-intensive of the three. Our tests showed that with $R=100$ randomly chosen nodes, the outcome of R-greedy is already very similar to FA-greedy.

\begin{figure}[htbp] 
	\centering
	\includegraphics[width=0.495\textwidth]{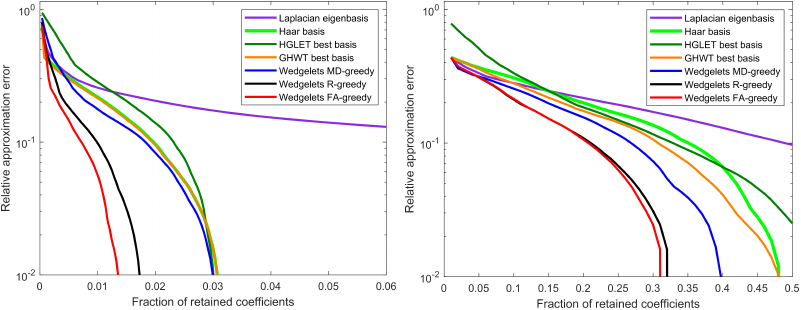} 	
	\vspace{-4mm}
	\caption{Comparison between best $m$-term approximation using BWP wavelets (FA-greedy, R-greedy with $R = 50$ and MD-greedy) and non-adaptive Haar-type wavelet dictionaries for the test functions $f_2$ (left) and $f_3$ (right).}
	\label{fig:BWP-comparison}
\end{figure}
	
In a second test, we compare the best $m$-term approximation of the adaptive BWP wavelets with four non-adaptive basis systems from the literature: a non-adaptive Haar basis, a Laplacian eigenbasis, a hierarchical graph Laplacian eigensystem (HGLET, \cite{IrionSaito2014}) and a generalized Haar-Walsh basis system (GHWT, \cite{IrionSaito2014b,IrionSaito2017}). The latter two are Haar-type dictionaries that contain, beside Haar functions, also Laplacian eigenfunctions on subgraphs (HGLETs) and generalized Walsh basis functions (GHWTs). For the comparison with the four non-adaptive basis systems we used the MTSG toolbox developed for the works \cite{IrionSaito2014,IrionSaito2014b,IrionSaito2017}. We can see that the data-adaptive FA-greedy and R-greedy schemes are able to generate sparser representations of the test functions in which a considerably smaller amount of the wavelet coefficients is required compared to the non-adaptive basis systems. While this higher sparseness is an aimed-at desirable property for data compression, adaptive basis systems require additional memory to store the respective geometric information. Therefore, when using adaptive BWPs for the compression of graph data also the additional storage costs of the BWPs as estimated in Section \ref{sec:storage} have to be taken into account.  

\begin{figure}[htbp] 
	\centering
	\includegraphics[width=0.495\textwidth]{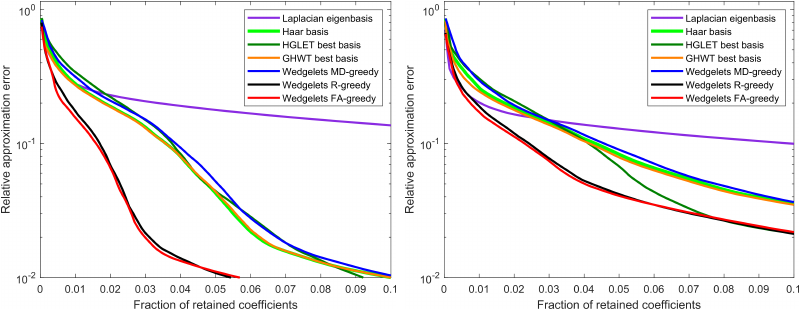} 	
	\vspace{-4mm}
	\caption{Comparison of best $m$-term approximations using Haar-type wavelet dictionaries for the piecewise smooth test function $f_4$ and gradient parameter $\alpha = 0.1$ (left) and $\alpha = 0.5$ (right).}
	\label{fig:BWP-comparison2}
\end{figure}

In a further test, we compare the same basis systems on the Minnesota graph for the approximation of piecewise smooth test functions. For this, we modify the piecewise constant signal $f_1$ by adding a gradient term, i.e., we consider the signal
\[f_4(\node{v}) = f_1(\node{v}) + \alpha(x_{\node{v}} - \bar{x}_V), \quad \alpha \in \Rr,\]
where $\bar{x}_V$ denotes the mean value of $x_{\node{v}}$ over $V$. As the parameter $\alpha$ increases, the function $f_4$ gets dominated by the gradient term and a dictionary of piecewise constant signals will struggle to approximate $f_4$ in a sparse way. This is confirmed by our simulation shown in Fig. \ref{fig:BWP-comparison2}. As $\alpha$ gets larger, the best $m$-term approximation errors using a pure Haar basis system (adaptive or non-adaptive) deteriorate. The difference between adaptive and non-adaptive basis systems is however still significant. It is also visible that basis systems that are better adapted to piecewise smooth functions as the HGLETs display an improved performance.      

\subsection{BWPs for the compression of 2D images}
\begin{figure*}[htbp]
	\centering
	\includegraphics[width=0.95\textwidth]{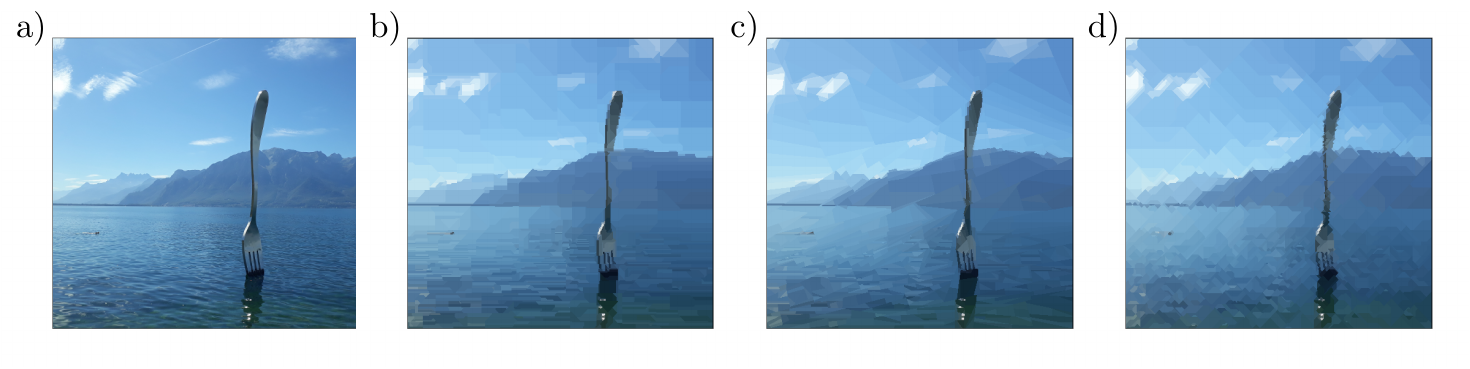}
	\vspace{-4mm}
	\caption{Role of the metric in BWP a) original image with $500 \times 451$ pixels; b)c)d) R-greedy compression with $M = 1000$ nodes, $R = 500$ as well as b) the $1$-norm, c) the $2$-norm and d) the infimum norm for the pixel distance. }
	\label{fig:BWP-fork}
\end{figure*}
As adaptive partitioning tools for discrete domains, BWPs can also be used for the piecewise approximation and compression of images. An image can be naturally thought of as a finite rectangular grid of pixels and interpreted as a graph. Pixels close to each other are therein linked by a weighted edge. The structure and the weights of the single edges determine the local dependencies in the image and have therefore a strong influence on the outcome of the greedy algorithms.

A simple qualitative comparison of the role of the used metric is given in Fig. \ref{fig:BWP-fork}. The $1$-norm and the infimum norm for the distance of the pixels lead to partitions with a rather rectangular or rhomboid wedge structure. On the other hand, wedges generated by the $2$-norm seem to be more anisotropic and slightly better adapted to the edges of the image.    

\begin{figure*}[htbp]
	\centering
	\includegraphics[width=0.95\textwidth]{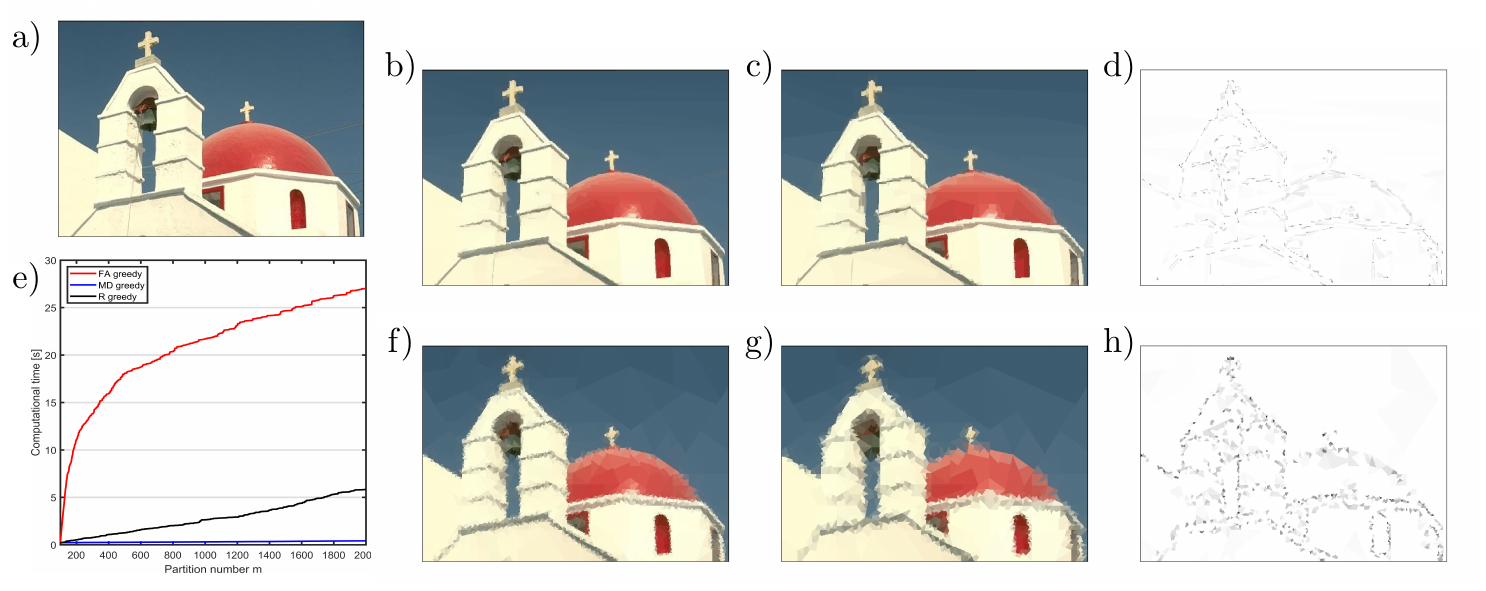} 
	\vspace{-4mm}
	\caption{BWP image compression. a) original $481 \times 321$-image; b)c) FA-greedy BWP compression for $M = 2000$, $M =1000$; d) wavelet details between b) and c); e) Computational times of the BWP variants; f)g) MD-greedy BWP compression for $M = 2000$, $M = 1000$; h) wavelet details between f) and g).}
	\label{fig:BWP-church}
\end{figure*}

\begin{figure*}[htbp]
	\centering
	\includegraphics[width=0.95\textwidth]{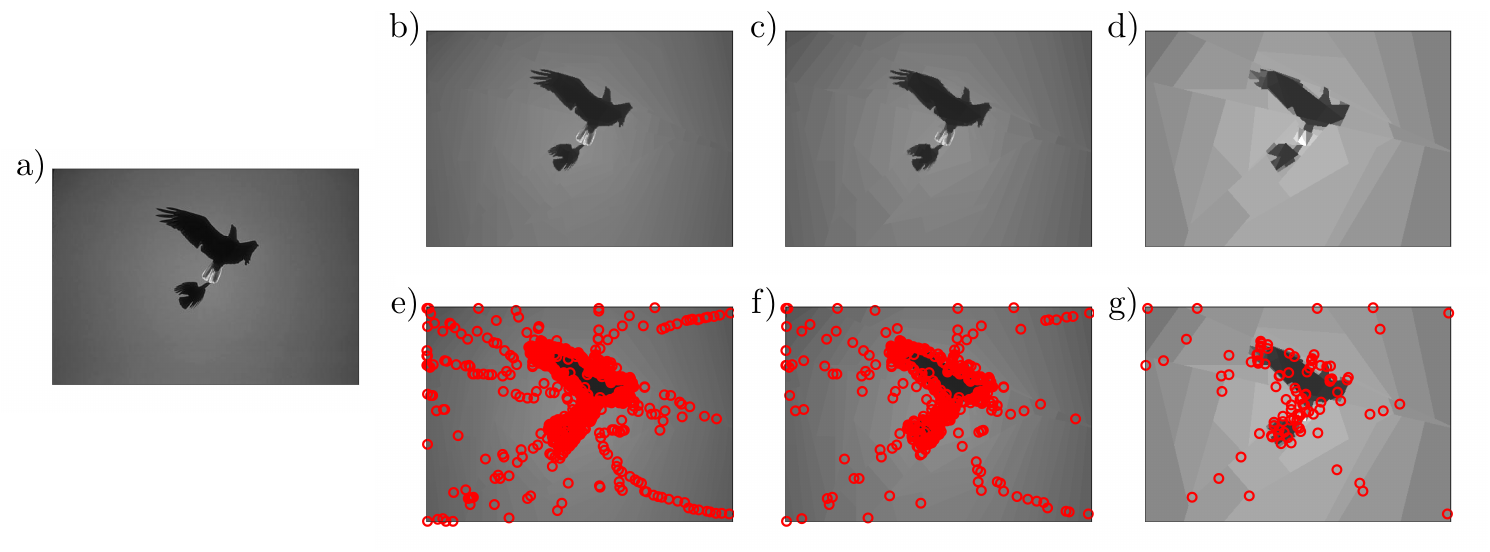} 
	\vspace{-4mm}
	\caption{BWP image encoding. a) original with $481 \times 321$ pixels; b)c)d) R-greedy compression with $1000$, $500$ and $100$ nodes, $R = 500$; e)f)g) respective node distributions for the approximations in b)c)d). The corresponding PSNR values are b) 40.762 dB, c) 37.935 dB, and d) 31.827 dB.}
	\label{fig:BWP-eagle}
\end{figure*}

We next compare the performance of the FA-greedy and the MD-greedy method for the compression of images. In the example given in Fig. \ref{fig:BWP-church}, we see that, as expected, the FA-greedy performs considerably better. In particular, the wavelet details in the FA-greedy scheme are smaller and more distinguished than for MD-greedy when using the same number of wedge splits. Regarding the distribution of the center nodes $Q$, we see further in Fig. \ref{fig:BWP-eagle} that the adaptive BWP scheme (in this case a R-greedy scheme) selects the new nodes increasingly closer to the edges of the image such that most refinements of the partitions are performed in those regions where the gradients are large. 

\begin{figure*}[htbp]
	\centering
	\includegraphics[width=0.95\textwidth]{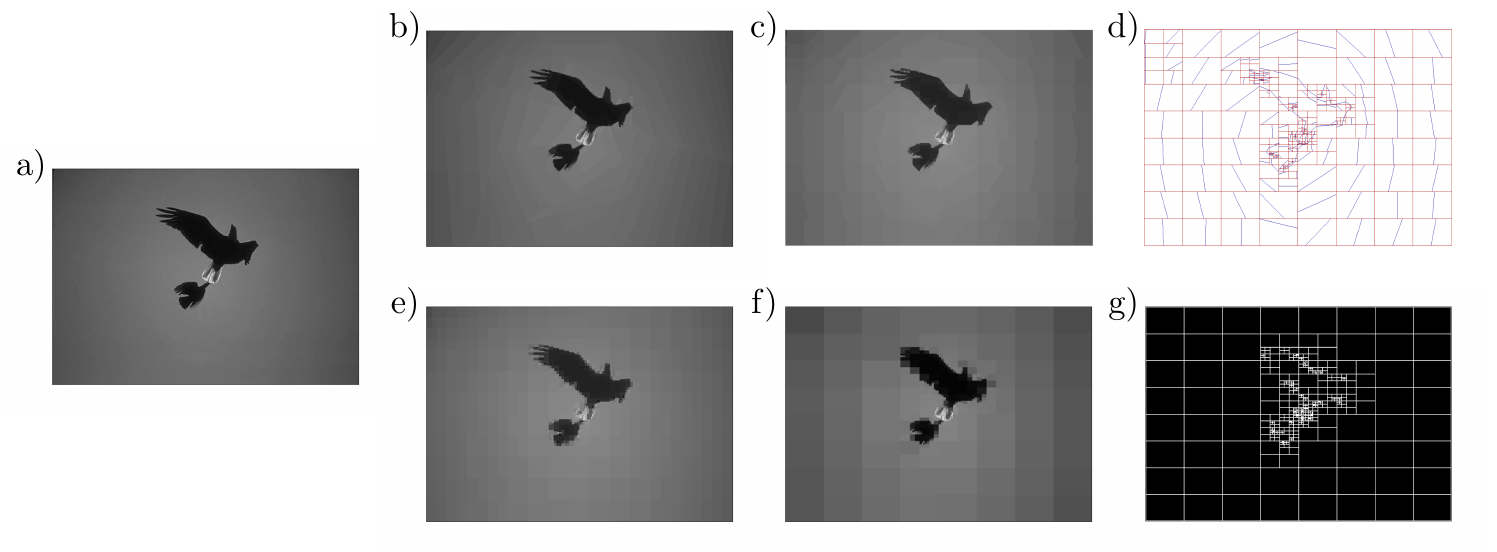} 
	\vspace{-4mm}
	\caption{Comparison of $4$ image compression techniques based on piecewise approximation: 
	a) original $481 \times 321$ image; b) graph wedgelet compression using $500$ most relevant BWP wavelet coefficients (PSNR: 38.297 dB) c)d) continuous wedgelet compression using $506$ wedges (PSNR: 36.828 dB) e) Haar wavelet compression using $500$ most relevant coefficients (PSNR: 34.764 dB) f)g)  quadtree compression with $505$ blocks (PSNR: 31.662 dB).}
	\label{fig:BWP-eagle-compare}
\end{figure*}

Finally, in Fig. \ref{fig:BWP-eagle-compare}, we compare the approximation quality of our adaptive BWP algorithm with three well-known segmentation based compression schemes from the literature. The first is a classical bivariate Haar wavelet transform in which the image is decomposed in uniform dyadic blocks and $6$ hierarchical levels. The most relevant wavelet coefficients of the image are then selected according to the Birg\'e-Massart strategy \cite{BirgeMassart1997}. In Fig. \ref{fig:BWP-eagle-compare} e) the respective compression for $500$ coefficients is illustrated. We compare this with a compression using the most relevant geometric wavelets in a R-greedy BWP tree ($R = 500$). Generating the BWP tree for $M = 4000$ and selecting the $500$ most relevant coefficient pairs $c_{(\node{q}_j,\node{q}_i)}^\pm$ provided by Algorithm \ref{alg:waveletencoding}, we obtain the compressed image in Fig. \ref{fig:BWP-eagle-compare} b). Qualitatively, the contours of the image are more pronounced for the adapted BWP wavelets, while in the classical Haar wavelet approach block artifacts are visible. A further indication for the higher image quality of BWP compression is the larger peak signal to noise ratio (PSNR).  

The two other compression methods are a continuous wedgelet decomposition (we use the implementation of \cite{Friedrich2007}) and a quadtree decomposition \cite{Samet1985} into adaptively generated dyadic blocks. The resulting image approximations shown in Fig. \ref{fig:BWP-eagle-compare} c)d)f)g) (using $506$ and $505$ segments, respectively) are compared with the graph wedgelet approximation in Fig. \ref{fig:BWP-eagle} c)f) (using $500$ wedgelets). It is visible that the number of quadtree blocks and continuous wedges is still quite low for a good resolution of the original image. This is also indicated by the lower PSNR of the resulting approximations. The graph wedgelets on the other hand display a higher adaptivity to the original image with a larger PSNR value. Our numerical tests also indicate that for smaller values of $M$ the generation of the wedgelet approximation alone, without thresholding the small geometric wavelet components, provides already good compression results. This is visible in the comparison between Fig \ref{fig:BWP-eagle} c) and \ref{fig:BWP-eagle-compare} b), where almost no difference between the two approximations is visible.

\section{Conclusion}
We introduced graph wedgelets: a novel type of geometric Haar-type basis functions on graphs that are able to efficiently capture the geometric information of signals using adaptive binary wedge splits that minimize a local $\mathcal{L}^2$-error. These discrete wedgelets are organized in terms of a binary wedge partitioning tree that can be encoded compactly in terms of a finite ordered sequence of graph nodes. We illustrated that these BWP trees provide promising dictionaries for the sparse representation of graph signals and can be applied to the compression of signals and images. From a theoretical point of view we showed that the geometric wavelets based on a near-best wedgelet partitioning tree provide quickly converging $m$-term approximants if the approximated signal is in a properly defined Besov-type smoothness class.  
 
\section*{Acknowledgment}
The author thanks the IN$\delta$AM research group GNCS, the Italian Research Network on Approximation (RITA), and the UMI-TAA research group for their support.

\section{Appendix}

\subsection{Proof of Theorem \ref{thm:BGPproperties}}

\begin{itemize}
 \item[(i)] As $\mathcal{T}$ is complete and binary, it contains $n$ leaves, and at least two of them are siblings (assuming that $n > 1$, for $n = 1$ the tree contains just a single root). Pruning the tree $\mathcal{T}$ by removing this pair of siblings, we obtain a reduced tree with $n-1$ leaves. An induction argument over the number $n$ of leaves therefore tells us that $ | \mathcal{T} | = 2n -1$.
 \item[(ii)] The completeness of $\mathcal{T}$ implies that we can decompose any signal $f$ in terms of the $n$ leaves of $\mathcal{T}$ as 
  $$f = \sum_{i = 1}^{n} f(\node{v}_i) \chi_{\{\node{v}_i\}} =  \sum_{W \in \mathcal{P}^{(n)}} \frac{\langle f, \chi_{W} \rangle}{|W|} \chi_{W}.$$
Now, by the definition of the wavelets $\psi_{W}(f)$ in \eqref{eq:geometricwavelet}, we recursively obtain
\begin{align*} f &= \sum_{W \in \mathcal{P}^{(n)} \setminus \mathcal{P}^{(n-1)}} \psi_{W}(f) + \sum_{W \in \mathcal{P}^{(n-1)}} \frac{\langle f, \chi_{W} \rangle}{|W|} \chi_{W}\\
&= \sum_{k = 2}^n \; \sum_{W \in \mathcal{P}^{(k)} \setminus \mathcal{P}^{(k-1)}} \!\!\!\!\!\!\!\!\! \psi_{W}(f) \; + \; \psi_V(f) = \sum_{W \in \mathcal{T}} \psi_{W}(f).
\end{align*}
Thus, using the ordered elements $W_i \in \mathcal{T}$, $i \in \{1, \ldots, 2n -1\}$ (according to the $\mathcal{L}^2$-norm of the wavelets), we obtain the representation in $(ii)$. 
\item[(iii)] In the proof of Theorem \ref{thm:Jackson}, we will show that for every BGP($\rho$) tree $\mathcal{T}$ the inequality 
\begin{align*}
\|f - \mathcal{S}_m(f)\|_{\mathcal{L}^2(V)} &\leq C m^{-\alpha} \mathcal{N}_r(f,\mathcal{T})
\end{align*}
holds true with a constant $C >0$ that depends only on $\rho$ (Theorem \ref{thm:Jackson} is actually formulated in terms of near-best BGP($\rho$) trees, the first part of the proof works however for all BGP($\rho$) trees). This implies that
\begin{align*}
\|f\|_{\mathcal{L}^2(V)} &\leq \|f - \mathcal{S}_1(f)\|_{\mathcal{L}^2(V)} + \|\mathcal{S}_1(f)\|_{\mathcal{L}^2(V)} \\ &\leq (C + 1)  \mathcal{N}_r(f,\mathcal{T}),
\end{align*}
and, thus, the statement of the theorem.

\end{itemize}

\subsection{Proof of Theorem \ref{thm:estimate}.}

\noindent We adapt the proof of \cite[Theorem 3.5]{DekelLeviatan2003}, in which a similar equivalence has been shown for binary space partitions in a continuous setting. To obtain the result of Theorem \ref{thm:estimate} for near best BGP$(\rho)$ trees, it suffices to show that for every BGP$(\rho)$ tree $\mathcal{T}$ the following equivalence holds:
\begin{equation} \ts \label{eq:auxequivalence} 
C_1 \mathcal{N}_r(f,\mathcal{T}) \leq \mathcal{M}_r(f,\mathcal{T}) \leq C_2 \mathcal{N}_r(f,\mathcal{T}),\end{equation}
where
\[\mathcal{M}_r(f,\mathcal{T}) = \left( \sum_{W \in \mathcal{T}} |W|^{-\alpha r} \sup_{\node{w} \in W} \sum_{\node{v} \in W} |f(\node{v}) - f(\node{w})|^r \right)^{\frac{1}{r}},\]
and $C_1,C_2 > 0$ are constants that depend only on $\rho$ and $r$. 
We start with the first inequality in \eqref{eq:auxequivalence}. If $W'$ is a child of $W$ in $\mathcal{T}$, we can estimate the $r$-norm $\|\psi_{W'}(f)\|_{\mathcal{L}^r(V)}$ as
\begin{align} \label{eq:a1} \| &\psi_{W'}(f)\|_{\mathcal{L}^r(V)} =  \left\| \textstyle \left( \frac{\langle f, \chi_{W'} \rangle}{|W'|} - \frac{\langle f, \chi_{W} \rangle}{|W|}\right) \! \chi_{W'}(\node{v}) \right\|_{\mathcal{L}^r(V)} \notag \\
&\leq C \left(\left\| (\bar{f}_{W'} - f) \chi_{W'} \right\|_{\mathcal{L}^r(V)} + \left\| (\bar{f}_{W} - f) \chi_{W'} \right\|_{\mathcal{L}^r(V)} \right) \notag \\
&= C \left(\left\| \bar{f}_{W'} - f  \right\|_{\mathcal{L}^r(W')} + \left\| \bar{f}_{W} - f \right\|_{\mathcal{L}^r(W')} \right) , 
\end{align}
where $\bar{f}_{W'}$ and $\bar{f}_{W}$ denote the means of $f$ over the sets $W'$ and $W$, respectively. The constant $C$ in this estimate depends only on $r$. The estimate \eqref{eq:a1} allows us to proceed as follows and to obtain the first of the two inequalities in \eqref{eq:auxequivalence}:
\begin{align*} \label{eq:a1} \textstyle
 \mathcal{N}_r(f,\mathcal{T}) & = \left( \sum_{W \in \mathcal{T}} 
\|\psi_{W}(f)\|_{\mathcal{L}^2(V)}^r \right)^{\frac{1}{r}} \\ &\leq \left( \sum_{W \in \mathcal{T}} |W|^{r/2 - 1}
\|\psi_{W}(f)\|_{\mathcal{L}^r(V)}^r \right)^{\frac{1}{r}} \\
& \leq C' \left( \sum_{W \in \mathcal{T}} |W|^{r/2 - 1}
\|\bar{f}_{W} - f\|_{\mathcal{L}^r(W)}^r \right)^{\frac{1}{r}} \\
&\leq C'' \left( \sum_{W \in \mathcal{T}} |W|^{r/2-2}
\sum_{\node{v} \in W} \|f(\node{v}) - f\|_{\mathcal{L}^r(W)}^r \right)^{\frac{1}{r}} \\
&\leq C'' \left( \sum_{W \in \mathcal{T}} |W|^{- \alpha r} \sup_{\node{w} \in W}
\sum_{\node{v} \in W} |f(\node{v}) - f(\node{w})|^r \right)^{\frac{1}{r}} \\ 
& = C'' \mathcal{M}_r(f,\mathcal{T}).
\end{align*}
Here, for the first inequality we used a general form of H\"older's inequality. For the second inequality, we combined the bound in \eqref{eq:a1} with the fact that a BGP$(\rho)$ tree is balanced resulting in a constant $C'$ that depends on $0 < \rho < 1$ and on $r$. Also the subsequent constant $C''$ depends only on $\rho$ and $r$. In the last inequality, the relation $1/r = \alpha + 1/2$ was included.  

We consider now the second inequality in \eqref{eq:auxequivalence}. Based on the expansion $f = \sum_{W \in \mathcal{T}} \psi_W(f)$, we get
\begin{align*}
 \sup_{\node{w} \in W} &
\sum_{\node{v} \in W} |f(\node{v}) - f(\node{w})|^r \\& = \sup_{\node{w} \in W}
\sum_{\node{v} \in W} |\sum_{W' \in \mathcal{T}, W' \subset W} \psi_{W'}(f)(\node{v}) - \psi_{W'}(f)(\node{w}) |^r \\
&\leq C \sum_{\node{v} \in W} \sum_{W' \in \mathcal{T}, W' \subset W} |\psi_{W'}(f)(\node{v}) |^r \\ &= C \sum_{W' \in \mathcal{T}, W' \subset W} \|\psi_{W'}(f)\|_{\mathcal{L}^r(W')}^r,
\end{align*}
with a constant $C$ that depends only on $r$. This provides for $\mathcal{M}_r(f,\mathcal{T})$ the bound
\begin{align*}
 \mathcal{M}_r& (f,\mathcal{T})^r  \leq C \sum_{W \in \mathcal{T}} |W|^{- \alpha r} \sum_{W' \in \mathcal{T}, W' \subset W} \|\psi_{W'}(f)\|_{\mathcal{L}^r(W')}^r \\
 & = C \sum_{W \in \mathcal{T}} \sum_{ W' \subset W} \left(\frac{|W'|}{|W|}\right)^{\alpha r} |W'|^{- \alpha r} \|\psi_{W'}(f)\|_{\mathcal{L}^r(W')}^r \\
  & = C \sum_{W' \in \mathcal{T}} |W'|^{- \alpha r} \|\psi_{W'}(f)\|_{\mathcal{L}^r(W')}^r \sum_{ W \supset W'} \left(\frac{|W'|}{|W|}\right)^{\alpha r}. 
\end{align*}
The balancedness of $\mathcal{T}$ implies that $|W'| \leq \rho |W|$ for every parent $W$ of $W'$, and thus $|W'| \leq \rho^k |W^{(k)}|$ for every parent $W^{(k)}$ in the $k$-th generation before $W'$. Thus,
\begin{equation} \label{eq:ancestors} 
\sum_{ W \supset W'} \left(\frac{|W'|}{|W|}\right)^{\alpha r} \leq \sum_{ k = 1}^{\infty} \rho^{k \alpha r} \leq \frac{ \rho^{\alpha r}}{1-\rho^{\alpha r}},
\end{equation}
and we can conclude that
\begin{align*}
 \mathcal{M}_r(f,\mathcal{T})^r &  \leq \frac{C \rho^{\alpha r}}{1-\rho^{\alpha r}} \sum_{W' \in \mathcal{T}} |W'|^{- \alpha r} \|\psi_{W'}(f)\|_{\mathcal{L}^r(W')}^r \\
 & \leq C' \sum_{W' \in \mathcal{T}} \|\psi_{W'}(f)\|_{\mathcal{L}^2(W')}^r = C' \mathcal{N}_r(f,\mathcal{T})^r,
\end{align*}
with a constant $C'$ that depends only on $\rho$ and $r$. This gives the second inequality in \eqref{eq:auxequivalence}.

\subsection{Proof of Theorem \ref{thm:Jackson}}

The subsequent proof is an adaption of the techniques developed for the derivation of \cite[Theorem 3.4]{KaraivanovPetrushev2003} in which a Jackson estimate for piecewise polynomial approximation on nested triangulations in $\mathbb{R}^2$ is formulated. The same techniques have been used in \cite[Theorem 3.6]{DekelLeviatan2003} to derive a respective Jackson estimate for geometric wavelets with respect to binary space partitionings of convex domains. \\

To subdivide the geometric wavelets of the tree $\mathcal{T}_r(f)$ into dyadic blocks, we define for $\mu \in \mathbb{Z}$ the index sets $$I_\mu\! := \textstyle \left\{  i \, \left| \, 
\frac{\mathcal{N}_r(f,\mathcal{T}_r(f))}{2^{\mu}} \leq \|\psi_{W_i}(f)\|_{\mathcal{L}^2(V)} < \frac{\mathcal{N}_r(f,\mathcal{T}_r(f))}{2^{\mu-1}} \right. \right\}. $$
For the union of these index sets, we get
\[\bigcup_{\nu \leq \mu} I_{\nu} = \left\{i \;| \; \|\psi_{W_i}(f)\|_{\mathcal{L}^2(V)} \geq 2^{-\mu} \mathcal{N}_r(f,\mathcal{T}_r(f))\right\}.\]
Thus, by Definition \ref{def:renergy} of the $r$-energy  $\mathcal{N}_r(f,\mathcal{T}_r(f))$, we have
\[ |I_{\mu}| \leq \sum_{\nu \leq \mu} | I_{\nu} | = \big| \bigcup_{\nu \leq \mu} I_{\nu} \big| \leq 2^{\mu r} .\]
Now, setting $m := \sum_{\nu \leq \mu} | I_{\nu} |$, we obtain the estimate
\begin{align*}
\|f - \mathcal{S}_m(f)\|_{\mathcal{L}^2(V)} & \leq \left\| \sum_{\nu > \mu} \sum_{i \in I_{\nu}} |\psi_{W_i}(f)|\right\|_{\mathcal{L}^2(V)} \\ &\leq \sum_{\nu > \mu} \left\| \sum_{i \in I_{\nu}}  |\psi_{W_i}(f)| \right\|_{\mathcal{L}^2(V)}.
\end{align*}
The wavelets $\psi_{W_i}(f)$ are constant on the sets $W_i$, thus $\|\psi_{W_i}(f)\|_{\mathcal{L}^2(V)} = \|\psi_{W_i}(f)\|_{\mathcal{L}^{\infty}(V)} |W_i|^{1/2}$. This together with the definition of the index sets $I_{\nu}$ gives the bound
\begin{align*}
\|f - \mathcal{S}_m(f)& \|_{\mathcal{L}^2(V)}  
\leq \sum_{\nu > \mu} \left\| \sum_{i \in I_{\nu}}  \|\psi_{W_i}(f)\|_{\mathcal{L}^{\infty}(V)} \chi_{W_i} \right\|_{\mathcal{L}^2(V)} \\
&= \sum_{\nu > \mu} \left\| \sum_{i \in I_{\nu}}  \frac{\|\psi_{W_i}(f)\|_{\mathcal{L}^{2}(V)}}{|W_i|^{1/2}} \chi_{W_i} \right\|_{\mathcal{L}^2(V)} \\
& \leq \sum_{\nu > \mu} 2^{-\nu+1} \mathcal{N}_r(f,\mathcal{T}_r(f))  \left\| \sum_{i \in I_{\nu}} \frac{\chi_{W_i}}{|W_i|^{1/2}}  \right\|_{\mathcal{L}^2(V)} \!\!\!\!\!.
\end{align*}
The balancedness of the BGP$(\rho)$ tree implies that the $\mathcal{L}^2$-norm in the last inequality can be bounded by
\begin{align*} 
\left\| \sum_{i \in I_{\nu}} \frac{\chi_{W_i}}{|W_i|^{1/2}}  \right\|_{\mathcal{L}^2(V)} & \hspace{-4mm}\leq \left( \! \sum_{i \in I_{\nu}} \sum_{\node{v} \in W_i} \frac{1}{|W_i|} \left(\sum_{ W \supseteq W_i} \!\!\frac{|W_i|^{1/2}}{|W|^{1/2}}\right)^{\!\!\!2}  \right)^{\!\frac12} \\
& \leq \left( \sum_{i \in I_{\nu}} \sum_{\node{v} \in W_i} \frac{1}{|W_i|} \frac{1}{1-\rho^{1/2}}  \right)^{\frac12} \\
&= \left(1-\rho^{1/2} \right)^{-\frac12} |I_{\nu}|^{1/2},
\end{align*}
where, in the second inequality, we bounded the interior sum similarly as in \eqref{eq:ancestors}. Combining the last two estimates and the previous bound on the complexity $|I_{\nu}|$, we can conclude that
\begin{align*}
\|f - \mathcal{S}_m(f) & \|_{\mathcal{L}^2(V)} \leq C \sum_{\nu > \mu} 2^{-\nu} \mathcal{N}_r(f,\mathcal{T}_r(f)) |I_{\nu}|^{1/2} \\ & \leq C \mathcal{N}_r(f,\mathcal{T}_r(f)) \sum_{\nu > \mu} 2^{-\nu(1-r/2)} \\
&= C \mathcal{N}_r(f,\mathcal{T}_r(f)) 2^{-\mu(1-r/2)} \\ & \leq C m^{-1/r+1/2}  \mathcal{N}_r(f,\mathcal{T}_r(f)) \\ &= C m^{-\alpha} \mathcal{N}_r(f,\mathcal{T}_r(f)),
\end{align*}
with $C = 2 (1-\rho^{1/2})^{-1/2}$.
Now, the bound of Theorem \ref{thm:Jackson} follows from the first inequality in Theorem \ref{thm:estimate}.


\begin{thebibliography}{1}


\bibitem{BirgeMassart1997}
{L. Birg\'e and P. Massart.}
\newblock ``From Model Selection to Adaptive Estimation,''
\newblock In {\em Festschrift for Lucien Le Cam: Research Papers in Probability and Statistics,} Springer, pp. 55--88, 1997.


\bibitem{Bremer2006}
{J.C. Bremer, R. Coifman, M. Maggioni, and A.D. Szlam.}
\newblock ``Diffusion wavelet packets,''
\newblock {\em Appl. Comput. Harmon. Anal.}, vol 21, no. 1, pp. 95--112, 2006. 

\bibitem{cavoretto2021}
{R. Cavoretto, A. De Rossi, and W. Erb.}
\newblock  ``Partition of Unity Methods for Signal Processing on Graphs,''
\newblock {\em J. Fourier Anal. Appl.}, vol. 27, Art. 66., 2021.
  
\bibitem{cheng2016}
{X. Cheng, X. Chen, and S. Mallat.}
\newblock  ``Deep Haar scattering networks,''
\newblock {\em Inf. Inference}, vol. 5, no. 2, pp. 105--33, 2016.

\bibitem{ChuiFilbirMhaskar2015}
{C.K. Chui, F. Filbir, and H.N. Mhaskar.} 
\newblock{``Representation of functions on big data: Graphs and trees,''}
\newblock {\em Appl. Comput. Harm. Anal.}, vol. 38, no. 3, pp. 489--509, 2015.
  
\bibitem{Chung}
{F.R.K. Chung.}
\newblock {``Spectral Graph Theory,''}
\newblock {\em AMS}, Providence, RI, 1997. 

\bibitem{CohenDynHechtMirebeau2012}
{A. Cohen, N. Dyn, F. Hecht, and J-M. Mirebeau.}
\newblock ``Adaptive multiresolution analysis based on anistropic triangulations,''
\newblock {\em Math. Comput.}, vol. 81, no. 278, pp. 789--810, 2012.

\bibitem{CoifmanGavish2011}
{R. Coifman and M. Gavish.}
\newblock ``Harmonic analysis of digital data bases,''
\newblock In {\em Wavelets and Multiscale analysis,} Springer, pp. 161--197, 2011.

\bibitem{CoifmanMaggioni2006}
{R. Coifman and M. Maggioni.}
\newblock ``Diffusion wavelets,''
\newblock {\em  Appl. Comput. Harmonic Anal.} vol. 21, pp. 53--94, 2006.

\bibitem{DekelLeviatan2003}
{S. Dekel and D. Leviatan.} 
\newblock{ ``Adaptive multivariate approximation using binary space partitions and geometric wavelets,''}
\newblock {\em SIAM J. Numer. Anal.}, vol. 43, no. 2, pp. 707--732, 2005.

\bibitem{DemaretDynIske2006}
{L. Demaret, N. Dyn, and A. Iske.}
\newblock {``Image compression by linear splines over adaptive triangulations,''}
\newblock {\em Signal Process.}, vol. 86, no. 7, pp. 1604--1616, 2006.

\bibitem{DemaretIske2006}
{L. Demaret and A. Iske.}
\newblock {\em ``Adaptive image approximation by linear splines over locally optimal delaunay triangulations,''}
\newblock {IEEE Signal Proc. Lett.}, vol. 13, no. 5, pp. 281--284, 2006.

\bibitem{DemaretIske2015}
{L. Demaret and A. Iske.}
\newblock {\em ``Optimal N-term approximation by linear splines over anisotropic Delaunay triangulations,''}
\newblock {Math. Comput.}, vol. 84, no. 293, pp. 1241--1264, 2015. 

\bibitem{DeVore1998}
{R. DeVore.} 
\newblock{``Nonlinear approximation,''}
\newblock {\em Acta Numer.}, vol. 7, pp. 51--150, 1998. 

\bibitem{Donoho1999}
{D.L. Donoho.}
\newblock ``Wedgelets: Nearly minimax estimation of edges,''
\newblock {\em Annals of Statistics}, vol. 27, no. 3, pp. 859--897, 1999.
  
\bibitem{Friedrich2007}
{F. Friedrich, L. Demaret, H. F\"uhr, and K. Wicker.}
\newblock {``Efficient Moment Computation over Polygonal Domains with an Application to Rapid Wedgelet Approximation,''}
\newblock {\em SIAM J. Sci. Comput.}, vol. 29, no. 2, pp. 842--863, 2007.

\bibitem{GavishNadlerCoifman2010}
{M. Gavish, B. Nadler, and R. Coifman.} 
\newblock{``Multiscale Wavelets on Trees, Graphs and High Dimensional Data: Theory and Applications to Semi Supervised Learning,''}
\newblock In {\em ICML'10: Proccedings of the 27th International Conference on Machine Learning}, pp. 367--374, 2010.

\bibitem{Gonzalez1985}
{T.F. Gonzalez.}
\newblock {``Clustering to minimize the maximum intercluster distance,''}
\newblock {\em Theoretical Computer Science}, vol. 38, pp. 293--306, 1985. 

\bibitem{GodsilRoyle2001}
{G. Godsil and G. Royle.}
\newblock ``Algebraic Graph Theory,''
\newblock {\em Springer}, New York, 2001.

\bibitem{Hammond2011}
{D.K. Hammond, P. Vandergheynst, and R. Gribonval.}
\newblock {``Wavelets on graphs via spectral graph theory,''}
\newblock {\em Appl. Comput. Harm. Anal.}, vol. 30, no. 2, pp. 129--150, 2011.

\bibitem{IrionSaito2014}
{J. Irion and N. Saito.} 
\newblock{``Hierarchical graph Laplacian eigen transforms,''}
\newblock {\em Jpn. SIAM Lett.}, vol. 6, pp. 21--24, 2014.

\bibitem{IrionSaito2014b}
{J. Irion and N. Saito.} 
\newblock{``The generalized Haar-Walsh transform,''}
\newblock in {\em Proc. 2014 IEEE Workshop on Statistical Signal Processing}, pp. 472--475, 2014.

\bibitem{IrionSaito2017}
{J. Irion and N. Saito.} 
\newblock{``Efficient Approximation and Denoising of Graph Signals Using the Multiscale Basis Dictionaries,''}
\newblock {\em IEEE Trans. Signal Inf. Process. Netw. }, vol. 3, no. 3, pp. 607--616, 2017.

\bibitem{Jansen2014}
{M. Jansen, G.P. Nason, and B.W. Silverman.} 
\newblock{``Multiscale methods for data on graphs and irregular multidimensional situations,''}
\newblock {\em Journal of the Royal Statistical Society}, vol. 71, no. 1, pp. 97--125, 2009. 

\bibitem{KaraivanovPetrushev2003}
{B. Karaivanov and P. Petrushev.} 
\newblock{``Nonlinear piecewise polynomial approximation beyond Besov spaces,''}
\newblock {\em Appl. Comput. Harm. Anal.}, vol. 15, no. 3, pp. 177--223, 2003.

\bibitem{Krommweh2010}
{J. Krommweh.} 
\newblock{``Tetrolet transform: A new adaptive Haar wavelet algorithm for sparse image representation,''}
\newblock {\em J. Vis. Commun. Image Represent.}, vol. 21, no. 4, pp. 364--374, 2010.

\bibitem{LeonardiKunt1985}
{R. Leonardi and M. Kunt.}
\newblock {``Adaptive split-and-merge for image analysis and coding,''}
\newblock In {\em Proc. SPIE}, vol. 594, 1985.

\bibitem{Murtagh2007}
{F. Murtagh.}
\newblock {``The Haar wavelet transform of a dendrogram,''}
\newblock {\em J. Classification}, vol. 24, pp. 3--32, 2007.

\bibitem{Ortega2018} 
{A. Ortega, P. Frossard, J. Kovačević, J.M.F. Moura and P Vandergheynst.}
\newblock {Graph Signal Processing: Overview, Challenges, and Applications.}
\newblock In {\em Proceedings of the IEEE}, vol. 106, no. 5, pp. 808--828, 2018.

\bibitem{NarangOrtega2012}
{S.K. Narang and A. Ortega.}
\newblock {``Perfect reconstruction two-channel wavelet filter-banks for graph structured data,''}
\newblock {\em IEEE Trans. on Sig.
Proc.}, vol. 60, no. 6, pp. 2786--2799, 2012.

\bibitem{RadhaLeonardiNaylorVetterli1990}
{H. Radha, R. Leonardi, B. Naylor, and M. Vetterli.} 
\newblock{``Image representation using binary space partitioning trees,''}
\newblock In {\em  Proc. SPIE}, vol. 1360, pp. 639--650, 1990.

\bibitem{RadhaVetterliLeonardi1996}
{H. Radha, M. Vetterli, and R. Leonardi.} 
\newblock{``Image compression using binary space partitioning trees,''}
\newblock {\em  IEEE Trans. Image Proc.}, vol. 5, no. 12, pp. 1610--1624, 1996.  

\bibitem{RossiAhmed2015}
{R.A. Rossi and N.K. Ahmed.}
\newblock {``The Network Data Repository with Interactive Graph Analytics and Visualization,''} 
\newblock In {\em Proc. of the Twenty-Ninth AAAI Conference on Artificial Intelligence}, {\em http://networkrepository.com}, 2015.

\bibitem{RustamovGuibas2013}
{R. Rustamov and L.J. Guibas.}
\newblock {``Wavelets on Graphs via Deep Learning,''} 
\newblock In {\em Advances in Neural Information Processing Systems}, vol. 26, Curran Associates Inc., 2013. 

\bibitem{Samet1985}
{H. Samet.} 
\newblock{``Data structures for quadtree approximation and compression,''}
\newblock {\em  Communications of the ACM}, vol. 28, no. 9, pp. 973--993, 1985.

\bibitem{Sharpnack2013}
{J. Sharpnack and A. Singh.} 
\newblock{``Near-optimal and computationally efficient detectors for weak and sparse graph-structured patterns,''}
\newblock {\em  2013 GlobalSIP}, pp. 443-446, 2013. 

\bibitem{ShenOrtega2010}
{G. Shen and A. Ortega.} 
\newblock{``Transform-based distributed data gathering,''}
\newblock {\em IEEE Trans. on Sig. Proc.}, vol. 58, no. 7, pp. 3802--3815, 2010.

\bibitem{shuman2012}
{D.I. Shuman, B. Ricaud, and P. Vandergheynst.}
\newblock {``A windowed graph {F}ourier transform,''}
\newblock in {\em Proc. 2012 IEEE Statistical Signal Processing Workshop (SSP)}, pp. 133--136, 2012.

\bibitem{shuman2013}
{D.I. Shuman, S.K. Narang, P. Frossard, A. Ortega, and P. Vandergheynst.}
\newblock {``The emerging field of signal processing on graphs: Extending high-dimensional data analysis to networks and other irregular domains,''} 
\newblock {\em IEEE Signal Proc. Mag.}, vol. 30, no. 3, pp. 83--98, 2013.

\bibitem{shuman2016}
{D.I. Shuman, B. Ricaud, and P. Vandergheynst.}
\newblock {``Vertex-frequency analysis on graphs,''} 
\newblock {\em Appl. Comput. Harm. Anal.}, vol. 40, no. 2, pp. 260--291, 2016.

\bibitem{shuman2020}
{D.I. Shuman.}
\newblock {``Localized spectral graph filter frames: A unifying framework, survey of design considerations, and numerical comparison,''}
\newblock {\em IEEE Sig. Proc. Mag}, vol. 37, no. 6, pp. 43--63, 2020. 

\bibitem{Szalam2005}
{A.D. Szlam, M. Maggioni, R.R. Coifman, and J.C. Bremer.}
\newblock {``Diffusion-driven multiscale analysis on manifolds and graphs: top-down and bottom-up constructions,''} 
\newblock In {\em Proc. SPIE}, vol. 5914, Wavelets XI, 59141D, 2005.

\bibitem{vonLuxburg2007}
{U. von Luxburg.}
\newblock {``A tutorial on spectral clustering,''}
\newblock {\em Stat. Comput.}, vol. 17, pp. 395--416, 2007.

\bibitem{WakinRomberg2003}
{M. Wakin, J. Romberg, H. Choi, and R. Baraniuk.}
\newblock {``Geometric methods for wavelet-based image compression,''}
\newblock In {\em Proc. SPIE}, vol. 5207, no. 1 \& 2, pp. 507--520, 2003. 


\end{thebibliography}
\end{document}